\documentclass{article}
\usepackage[utf8]{inputenc}

\usepackage[colorlinks=true, linkcolor=blue, citecolor=blue, urlcolor=blue]{hyperref}

\usepackage[margin=2cm]{geometry}

\usepackage{authblk}
\usepackage{amsmath}
\usepackage{graphics,graphicx}
\usepackage{amssymb}
\usepackage{multicol}
\usepackage{tabularx}
\usepackage[labelfont=bf]{caption}
\newcommand{\nat}{Nature}  
  
\newcommand{\apj}{Astrophys.~J.}      
\newcommand{\aj}{Astron.~J.}      
\newcommand{\aaps}{Astron.~Astrophys.~Supp.}  
\newcommand{\aap}{Astron.~Astrophys.}  
\newcommand{\aapr}{Astron.~Astrophys. Rev.}  
\newcommand{\apjl}{Astrophys.~J.~Letters}   
\newcommand{\mnras}{Mon.~Not.~R.~Astron.~Soc.} 

\newcommand{\pasp}{Pub.~Astron.~Soc.~Pacific}   
\newcommand{\araa}{Ann.~Rev.~Astron.~Astrophys} 

\newcommand{\procspie}{Proc. SPIE}
\newcommand{\pasa}{Publ.~Astron.~Soc.~Australia}
\newcommand{\ssr}{Space~Sci.~Rev.}
\newcommand{\baas}{Bull.~American~Astron.~Soc.}

\usepackage[nolist,nohyperlinks]{acronym}

\usepackage{siunitx}
\usepackage{changepage}
\newcommand{\difx}{DiFX}
\newcommand{\polconvert}{PolConvert}
\begin{acronym}
\acro{sn}[S/N]{signal-to-noise ratio}
\acro{vlbi}[VLBI]{very long baseline interferometry}
\acroplural{vlbi}[VLBI]{Very-long-baseline interferometry}
\acro{grmhd}[GRMHD]{general relativistic magnetohydrodynamics}
\acro{mhd}[MHD]{magnetohydrodynamics}
\acro{as}[as]{arcseconds}
\acro{agn}[AGN]{Active Galactic Nuclei}
\acro{llagn}[LLAGN]{low-luminosity AGN}
\acro{cena}[Cen~A]{Centaurus~A}
\acroplural{cena}[Cen~A, NGC~5128]{Centaurus~A}
\acro{sgra}[Sgr\,A*]{Sagittarius~A*}
\acro{agn}[AGN]{active galactic nuclei}
\acro{eht}[EHT]{Event Horizon Telescope}
\acro{bhc}[BHC]{\href{https://blackholecam.org}{BlackHoleCam}}
\acro{tanami}[TANAMI]{Tracking Active Galactic Nuclei with Austral Milliarcsecond Interferometry}
\acro{smbh}[SMBH]{supermassive black hole}
\acroplural{smbh}[SMBHs]{supermassive black holes}
\acro{aa}[ALMA]{Atacama Large Millimeter/submillimeter Array}
\acro{ap}[APEX]{Atacama Pathfinder Experiment}
\acro{pv}[PV]{IRAM~30\,m Telescope}
\acro{jc}[JCMT]{James Clerk Maxwell Telescope}
\acro{lm}[LMT]{Large Millimeter Telescope Alfonso Serrano}
\acro{sp}[SPT]{South Pole Telescope}
\acro{sm}[SMA]{Submillimeter Array}
\acro{az}[SMT]{Submillimeter Telescope}
\acro{c}[c]{speed of light}
\acro{pc}[pc]{parsec}
\acro{gr}[GR]{general relativity}
\acro{aips}[\textsc{aips}]{\href{http://www.aips.nrao.edu}{Astronomical Image Processing System}}
\acro{casa}[\textsc{casa}]{\href{https://casa.nrao.edu}{Common Astronomy Software Applications}}
\acro{symba}[\textsc{symba}]{\href{https://bitbucket.org/M_Janssen/symba}{SYnthetic Measurement creator for long Baseline Arrays}}
\acro{rpicard}[\textsc{Rpicard}]{\href{https://bitbucket.org/M_Janssen/picard}{Radboud PIpeline for the Calibration of high Angular Resolution Data}}
\acro{hops}[\textsc{hops}]{\href{https://www.haystack.mit.edu/tech/vlbi/hops.html}{Haystack Observatory Postprocessing System}}
\acro{hbt}[HBT]{Hanbury Brown and Twiss}
\acro{tov}[TOV]{Tolman-Oppenheimer-Volkoff}
\acro{mri}[MRI]{magnetorotational instability}
\acro{adaf}[ADAF]{advection-dominated accretion flow}
\acro{adios}[ADIOS]{adiabatic inflow-outflow solution}
\acro{cdaf}[CDAF]{convection-dominated accretion flow}
\acro{bz}[BZ]{Blandford-Znajek}
\acro{bp}[BP]{Blandford-Payne}
\acro{em}[EM]{electromagnetic}
\acro{blr}[BLR]{broad-line region}
\acro{nlr}[NLR]{narrow-line region}
\acro{ism}[ISM]{interstellar medium}
\acro{edf}[eDF]{electron distribution function}
\acro{pic}[PIC]{particle-in-cell}
\acro{sane}[SANE]{standard and normal evolution}
\acro{mad}[MAD]{magnetically arrested disk}
\acro{jive}[JIVE]{\href{http://www.jive.eu}{Joint Institute for VLBI ERIC}}
\acro{nrao}[NRAO]{\href{https://www.nrao.edu}{National Radio Astronomy Observatory}}

\acro{muas}[$\mu$as]{microarcseconds}
\acroplural{agn}[AGN]{Active galactic nuclei}
\acro{jy}[Jy]{jansky}
\acro{pa}[PA]{position angle}
\acro{srmhd}[SRMHD]{special relativistic magnetohydrodynamics}
\end{acronym}

\title{Event Horizon Telescope observations of the jet launching and collimation in Centaurus A}

\author[1,2*]{Michael Janssen}
\author[2]{Heino Falcke}

\author[3]{Matthias Kadler}

\author[1]{Eduardo Ros}

\author[4,5]{Maciek Wielgus}

\author[6,7,4]{Kazunori Akiyama}

\author[8.9]{Mislav Balokovi\'c}

\author[4,5]{Lindy Blackburn}

\author[4,5,10]{Katherine L. Bouman}

\author[11,12]{Andrew Chael}

% ...

\affil[1]{Max-Planck-Institut f\"ur Radioastronomie, Auf dem H\"ugel 69, D-53121 Bonn, Germany}

\affil[2]{Department of Astrophysics, Institute for Mathematics, Astrophysics and Particle Physics (IMAPP), Radboud University, P.O. Box 9010, 6500 GL Nijmegen, The Netherlands}

\affil[3]{Lehrstuhl f\"{u}r Astronomie, Universit\"{a}t W\"{u}rzburg, Emil-Fischer Str. 31, 97074 W\"{u}rzburg, Germany}

\affil[4]{Black Hole Initiative at Harvard University, 20 Garden Street, Cambridge, MA 02138, USA}

\affil[5]{Center for Astrophysics $\mid$ Harvard \& Smithsonian, 60 Garden Street, Cambridge, MA 02138, USA}

\affil[6]{Massachusetts Institute of Technology Haystack Observatory, 99 Millstone Road, Westford, MA 01886, USA}

\affil[7]{National Astronomical Observatory of Japan, 2-21-1 Osawa, Mitaka, Tokyo 181-8588, Japan}

\affil[8]{Yale Center for Astronomy \& Astrophysics, 52 Hillhouse Avenue, New Haven, CT 06511, USA}

\affil[9]{Department of Physics, Yale University, P.O. Box 2018120, New Haven, CT 06520, USA}

\affil[10]{California Institute of Technology, 1200 East California Boulevard, Pasadena, CA 91125, USA}

\affil[11]{Princeton Center for Theoretical Science, Jadwin Hall, Princeton University, Princeton, NJ 08544, USA}

\affil[12]{NASA Hubble Fellowship Program, Einstein Fellow}

% ...
\affil[*]{Correpsonding author. Email: \href{mailto:mjanssen@mpifr-bonn.mpg.de}{mjanssen@mpifr-bonn.mpg.de}}
\affil[(**)]{The Event Horizon Telescope Collaboration}
\date{}                     
\usepackage[super,sort,comma,numbers]{natbib}
\usepackage{graphicx}

\begin{document}

\maketitle

\section*{}
\newpage
\textbf{\acp{vlbi} observations of active galactic nuclei at millimeter wavelengths have the power to reveal the launching and initial collimation region of extragalactic radio jets, down to 10\,--\,100 gravitational radii ($r_g \equiv G M /c^2$) scales in nearby sources \citep{2017Boccardi}.
Centaurus~A is the closest radio-loud source to Earth \citep{Harris2010}. It bridges the gap in mass and accretion rate between the \acp{smbh} in Messier\,87 and our galactic center. 
A large southern declination of \ang{-43} has however prevented \ac{vlbi} imaging of Centaurus~A below $\lambda$\,1\,cm thus far.
Here, we show the millimeter \ac{vlbi} image of the source, which we obtained with the Event Horizon Telescope at 228\,GHz.
Compared to previous observations \citep{Mueller2014}, we image Centaurus~A's jet at a tenfold higher frequency and sixteen times sharper resolution and thereby probe sub-lightday structures.
We reveal a highly-collimated, asymmetrically edge-brightened jet as well as the fainter counterjet.
We find that Centaurus~A's source structure resembles the jet in Messier\,87 on $\sim500\,r_g$ scales remarkably well.
Furthermore, we identify the location of Centaurus~A's \ac{smbh} with respect to its resolved jet core at $\lambda$\,1.3\,mm and conclude that the source's event horizon shadow \citep{2000Falcke} should be visible at THz frequencies. This location further supports the universal scale invariance of black holes over a wide range of masses \citep{2003Merloni,2004Falcke}.}

In this work, we present the first image of \ac{cena} obtained by the \ac{eht} with a nominal resolution of 25\,\ac{muas} at $\lambda$\,1.3\,mm.
For a black hole mass of \mbox{$\left(5.5 \pm 3\right) \times 10^7\,M_\odot$} \citep{Neumayer2010}, we are probing jet structures down to scales of $\sim 200\,r_g \approx 0.6$ light days. It has recently become possible to model these scales with sophisticated \ac{grmhd} simulations \citep{hamr2}, where jet ejection and their symbiotic relationship with accretion flows are simulated from first principles.
We have observed \ac{cena} in a six-hour-long track on April 10, 2017.
The \ac{eht}, as a novel and heterogeneous high-frequency \ac{vlbi} array, poses unique calibration challenges.
To obtain robust results, independent of assumptions made during the data calibration, we base our scientific analysis on two data sets, which we obtained from two independent calibration pipelines: rPICARD \citep{Janssen2019} and EHT-HOPS \citep{Blackburn2019} (Methods~\ref{sec:methods:pipelines}).

\begin{figure}[h!]
\centering
\includegraphics[width=0.85\textwidth]{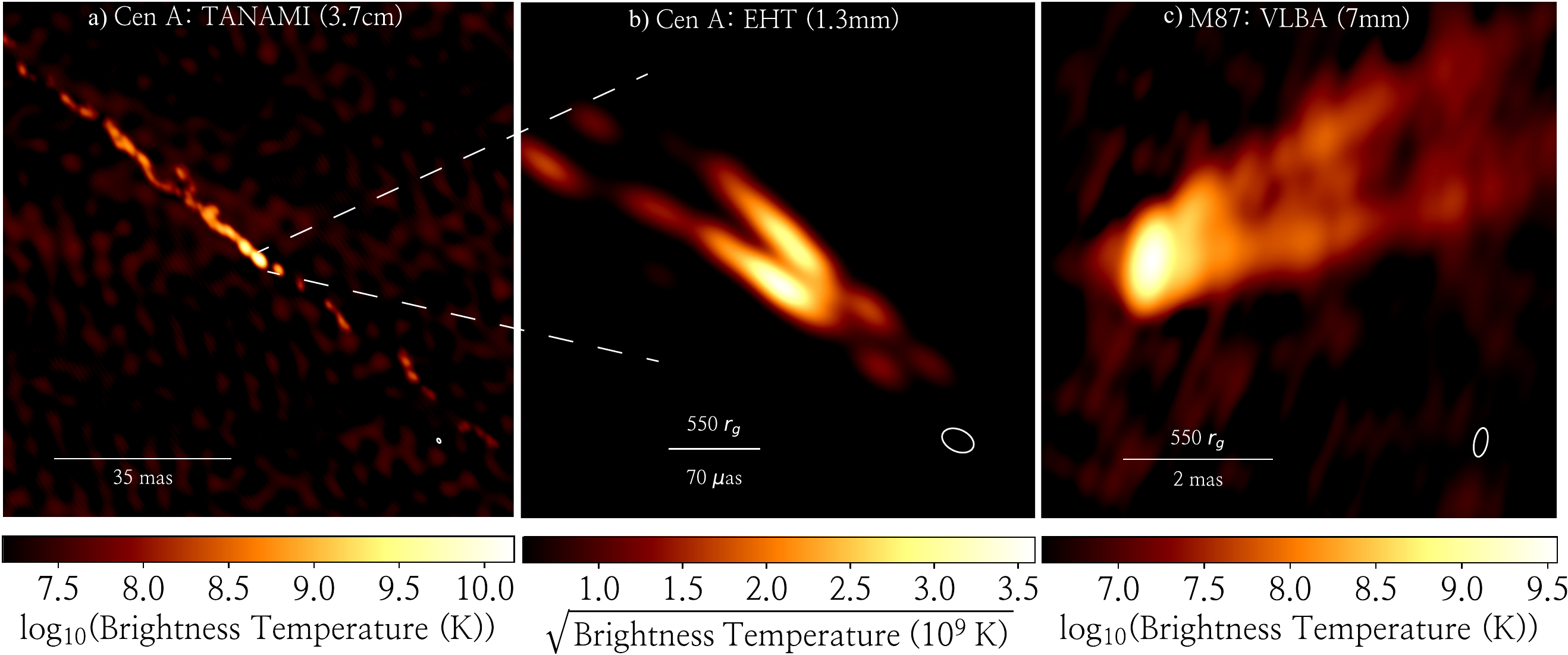} 
\caption{\textbf{The jet structure of \ac{cena} in comparison with M\,87.}
         The left panel (a) shows the large scale jet of \ac{cena} from an 8\,GHz ($\lambda\,3.7$\,cm) TANAMI \citep{Ojha2010} observation in November 2011 \citep{Mueller2014} on a logarithmic color scale.
         The middle panel (b) depicts our final \ac{eht} image from April 2017, blurred to the nominal resolution for a uniform weighting of the visibilities (the beams are shown in the bottom right corners).
         The reconstruction is based on the rPICARD data and is shown on a square-root scale, where values below a brightness temperature of $3\times10^8$\,K are clipped, due to a lower dynamic range compared to the longer-wavelength observations. An unclipped and unconvolved version of this image is shown in \autoref{CenA-fig:im-analysis}.
         The right panel (c) shows the M\,87 jet at 43\,GHz ($\lambda\,7$\,mm) from a Very Long Baseline Array observation in June 2013 \citep{Walker2018,Janssen2019} on a logarithmic scale.
         North is up and east is to the left.
         The physical, linear scales of the full field of views shown in the three images are 2\,\ac{pc} for TANAMI (left), 0.007\,\ac{pc} for the EHT (middle), and 0.6\,\ac{pc} for the VLBA (right).
        }
\label{CenA-fig:data-model}
\end{figure}

\autoref{CenA-fig:data-model} presents our reconstruction of the jet image structure derived from the \ac{eht} data using a regularized maximum-likelihood method, next to the large-scale source morphology and the similarly edge-brightened morphology of the M\,87 jet on comparable gravitational scales.
These images are convolved with Gaussian beams set by their respective nominal instrumental resolutions, as per standard practice in radio-interferometric imaging, to suppress possibly spurious fine-scale structures in the image model.
The brightness temperatures $\mathcal{T}$\,[K] shown are related to flux densities $S$ in \ac{jy} through the observing wavelength $\lambda$, Boltzmann constant $k_\mathrm{B}$, and angular resolution element $\Omega$ as $\mathcal{T} = \lambda^2 \left(2 k_\mathrm{B} \Omega\right)^{-1} S$.
The $\lambda$\,1.3\,mm \ac{cena} jet has a narrow, collimated profile and exhibits one-sidedness, pronounced edge-brightening, and a northwest-southeast (NW-SE) brightness asymmetry.
The approaching jet extends towards the north-east and the faint counterjet is directed south-westwards.
The total compact flux density in our image is $\sim 2$\,\ac{jy}.
The identification of the jet apex and black hole position (Methods~\ref{sec:methods:apexpos}) is shown in the unconvolved image model of \autoref{CenA-fig:im-analysis}. We can use interferometer data with a high signal-to-noise ratio to super-resolve image features beyond the nominal resolution of the instrument. We therefore base our analysis on the robust features of the unconvolved image model.
We have verified the robustness of the counterjet feature with synthetic data studies (Supplementary~Fig.~\ref{CenA-SuppFIG:symba}). The estimated jet \ac{pa} on the sky of $\ang{48} \pm \ang{5}$ agrees with cm-wave VLBI observations \citep{Mueller2014}.
The cm-band data also constrain the inclination angle of the jet axis with respect to our line of sight to $\theta \sim \ang{12}$\,--\,\ang{45}, assuming the jet does not bend along the line of sight. 


The \ac{cena} $\lambda$\,1.3\,mm jet exhibits three types of brightness asymmetries ($\mathcal{R}$): between the jet and counterjet, the sheath and spine, and the NW vs. SE ridgelines (Methods~\ref{sec:methods:asymm}). We take the two bright radiating streams of the approaching- and counter-jet as jet `arms' and denote the maximum intensity region along each arm as `ridgeline'.
The jet-to-counterjet intensity ratio $\mathcal{R}_\mathrm{j/cj}$ can naturally be explained for a relativistic outflow with $\theta \neq \ang{90}$, where jet emission will be Doppler boosted and counterjet emission de-boosted. 
We find $\mathcal{R}_\mathrm{j/cj} \gtrsim 5$, which is in agreement with cm-wave \ac{vlbi} observations \citep{Mueller2014} and suggests that the initial acceleration of the jet occurs within the inner collimation region imaged in this study.

There is no jet spine emission in our image.
With synthetic data studies, we found that spine emission exceeding $\sim 20\,\%$ of the sheath radiation intensity would be detectable, i.e., $\mathcal{R}_\mathrm{sh/sp} > 5$ (Methods~\ref{sec:symba-tests}).
The intensities of the brightest, central SE and NW jet components in the unconvolved image are $\left( 32 \pm 8 \right) \times 10^9$\,K and $\left( 20 \pm 4 \right) \times 10^9$\,K, respectively.
The brightness ratio between these components follows as $\mathcal{R}_\mathrm{s/n} = 1.6 \pm 0.5$.


The collimation profile of the jet width $W$ follows a narrow expansion profile with distance to the apex $z$ as $W \propto z^k$ with $k=0.33 \pm 0.05{\mid}_{\mathrm{stat}} \pm 0.06{\mid}_{\mathrm{sys}}$ (\autoref{CenA-fig:jet_fit}).
Resolution and potentially optical depth effects prevent us from pinning down the jet opening angle $\psi_\mathrm{jet}$ at small $z$, where the jet converges towards the apex.
We denote the boundary between the inner convergence region and the outer jet with a clearly defined collimation and easily traceable jet ridgelines as $z_\mathrm{col}$.
For the brighter and straighter SE arm, we have $W(z_\mathrm{col} \approx 32\,\ac{muas}) \approx 25\,\ac{muas}$, i.e., the brightest jet component marks the boundary between the convergence and strongly collimated regions here (\autoref{CenA-fig:im-analysis}).
If we assume the two jet ridgelines to meet at the apex, we find $\psi_\mathrm{jet} \gtrsim \ang{40}$ as a conservative estimate.
Factoring in the range of possible $\theta$ values yields $\psi_\mathrm{int} \gtrsim \ang{10}$\,--\,\ang{30} for the intrinsic, deprojected opening angle (Methods~\ref{sec:methods:col}).


The M\,87 \citep{2018KimM87} (NGC~4486, 3C~274, Virgo~A), Markarian\,501 \citep{2009Piner}, and restarted 3C\,84 jets \citep{2018Giovannini} also show strong edge-brightening and large initial opening angles on comparable scales seen at similar inclination angles of $\sim \ang{18}$.
The expansion profile of \ac{cena} lies in between the parabolic profile of M\,87 ($k = 0.5$) and the almost cylindrical profile of 3C\,84 ($k = 0.2$), which implies a strong confinement of the 3C\,84 jet by a shallow pressure gradient from the ambient medium.
For the inner \ac{cena} jet, this suggests strong magnetic collimation or the presence of external pressure and density gradients of $P_\mathrm{ext} \propto z^{-4k} = z^{-1.3}$ and $\rho_\mathrm{ext} \propto z^{1-4k} = z^{-0.3}$ (Methods~\ref{sec:methods:amb}). 
Radiatively inefficient accretion flows alone, which are expected to operate in the M\,87, 3C\,84, and \ac{cena} sub-Eddington \ac{llagn} sources, have comparatively steeper pressure and density gradients \citep{1998Narayan}.
This may indicate the presence of winds, which are likely to be launched by this type of accretion flow.
The noticeable similarity and prominence of edge-brightened jet emission in M\,87, 3C\,84, and \ac{cena} suggests the dominance of jet sheath emission to be an emerging feature in \ac{llagn}.
In \ac{grmhd} simulations, the sheath manifests itself as interaction region between an accretion-powered outflow \citep{1982Blandford} and the fast jet spine, which is potentially powered by the black hole spin \citep{1977Blandford}. The mass-loaded sheath has a higher intrinsic emissivity compared to the evacuated spine.
The same type of \ac{llagn}-applicable \ac{grmhd} simulations also self-consistently develop a collimating helical magnetic field structure in the jet, which is confirmed observationally in many \ac{agn} \citep{2018Gabuzda}.
The dominating sheath emissivity and helical magnetic field structure provides a natural intrinsic explanation for the prevailing edge-brightening in \ac{llagn} and can also explain the NW-SE brightness asymmetry.
This model and alternative geometric explanations for the brightness asymmetries are discussed in the Methods~\ref{sec:methods:asymm} section.


The basic radiative properties of these jets can be analytically understood with a simple model \citep{1979Blandford}, where particle and magnetic energy density equipartition is assumed, while the particle density decays with $z^{-2}$. Under these conditions, an optically thick and self-absorbed compact feature is expected (the core), whose position $z_\mathrm{core}$ along the jet is frequency dependent with $z_\mathrm{core} \propto \nu^{-1}$ \citep{1995Falcke,1998Lobanov}. This radio core corresponds to the photosphere, where the optical depth $\tau(\nu)$ to photons at the observing frequency $\nu$ is unity. The jet is optically thick upstream and optically thin downstream. The photosphere moves closer to the jet apex at higher frequencies, until the point where either the launching point is reached near the horizon, or particle acceleration has not yet begun \citep{2017Romero}.  The scale of a jet `nozzle' emission cannot be smaller than the $\sqrt{27}\,r_g$ photon capture radius (Methods~\ref{sec:methods:bhloc}).

The combination of all emission regions along the jet gives rise to a flat to inverted radio spectrum, peaking at a maximum frequency $\widetilde{\nu}$, determined by the black hole mass $M$ and accretion rate $\dot M$ and scaling as $\widetilde{\nu} \propto M^{-1} \dot{M}^{2/3} \propto M^{-1} F_{\rm r}^{8/17} D^{16/17} \, $ \citep{1995Falcke,2003Heinz,2004Falcke}. Here, $D$ is the distance of the black hole to the observer and $F_{\rm r}$ the observed radio flux density. These scaling relations follow from the assumption that the jet's internal gas and magnetic pressures are linearly coupled to the accretion rate and maintain a fixed ratio along the jets. The proportionality constant between $\dot{M}$ and $F_{\rm r}$ generally depends on the jet's velocity, electron and magnetic energy densities, particle distribution spectrum, and inclination angle. Therefore, we are only able to make a first order estimates.
It should further be noted that X-ray binary observations \citep{2020Lucchini} have revealed a more complex relationship between $\widetilde{\nu}$ and $\dot{M}$, where the innermost particle acceleration zone in the jet may not remain stationary and source-specific accretion disk parameters come into play.
The same effects are expected to also influence $\widetilde{\nu}$ in \ac{agn}, which substantiates the fact that only order of magnitude estimates can be provided for $\widetilde{\nu}$.
We assume the brightest features in our image to correspond to the radio cores at 230\,GHz, which is discussed in Methods~\ref{sec:methods:nocore}.
Our assumption is affirmed by three consistent and independent measurements of $\widetilde{\nu}$, but future spectral information is needed for a definitive confirmation.
We show that $\widetilde{\nu}$ lies in the THz regime for \ac{cena} based on the core shift that we can determine from our image, scaling relations with the M\,87 jet, and the spectral energy distribution of \ac{cena}.

We take the distance from the brightest pixel in the image to the estimated position of the jet apex and obtain a core shift of $z_\mathrm{core} = 32 \pm 11$\,\ac{muas}.
Based on this distance and the uncertain inclination angle, we estimate that an observing frequency of $\widetilde{\nu}_\mathrm{CenA} \sim 10$\,--\,60\,THz (Methods~\ref{sec:methods:bhloc}) will reach the base of the jet at the black hole innermost stable circular orbit (ISCO). A caveat is that  we do not take the effect of the uncertain ambient medium into account in this simple picture.

Independently, we can use the above scaling relations to estimate the order of magnitude of $\widetilde{\nu}_\mathrm{CenA}$ by comparing the \ac{cena} jet with M\,87, which has $\widetilde{\nu}_\mathrm{M87} = 228$\,GHz \citep{Hada2011,eht-paperI}. 
For the cm jet radio core, a flux density of $\sim 1$\,\ac{jy} is measured for both sources \citep{Mueller2014,Walker2018}, which yields $\dot{M}_\mathrm{CenA} \sim 0.1 \dot{M}_\mathrm{M87}$ for the accretion onto the black hole and therefore \mbox{$\widetilde{\nu}_\mathrm{CenA} \sim 26 \, \widetilde{\nu}_\mathrm{M87} \approx 6$\,THz} (Methods~\ref{sec:methods:bhloc}), in agreement with our observations and the assumed position of the black hole at the jet apex within an order of magnitude. Based on comparable jet velocities \mbox{($\sim 0.3$\,c\,--\,0.5\,c)} and inclination angles ($\sim$\,\ang{20}), we have assumed the amount of Doppler boosting to be similar in both jets. 
The relation of accretion rates would constrain $\dot{M}_\mathrm{CenA}$ to be \mbox{$\lesssim 9 \times 10^{-5}\,M_\odot\,\mathrm{yr}^{-1}$ \citep{2014Kuo}} or $7 \times 10^{-5}\,\dot{M}_\mathrm{Edd}$ in terms of the Eddington accretion rate for an assumed radiative efficiency of 10\,\%.

The core spectral energy distribution of \ac{cena} peaks at $\sim 10^{13}$\,Hz \citep{Abdo2010}, which may be the equivalent of the submillimeter bump seen in Sgr\,A* \citep{1998Falcke,2019Bower}, and would further support our hypothesis.

Observed correlations between the masses of accreting black holes and their X-ray and radio luminosities form the basis of a unified fundamental plane of scale-invariant black hole accretion.
This scale-invariance has been derived based on stellar-mass black holes, which have a break frequency $\widetilde{\nu}_\mathrm{XRB}$ in the near-infrared, and supermassive $10^8\,M_\odot$\,--\,$10^{10}\,M_\odot$ \ac{agn}, where $\widetilde{\nu}_\mathrm{SMBH}$ lies in the radio to submillimeter regime \citep{2003Merloni,2004Falcke}.
With our observation, we demonstrate that the simple fundamental relations for the black hole jet activity still holds for a source with a mass of $5.5 \times 10^7\,M_\odot$ and $\widetilde{\nu}_\mathrm{CenA}$ in the THz regime, in between those two types of black holes.
Our method used to determine the optimal frequency to observe black hole shadows based on core shift, jet power, and source spectrum is in principle applicable to any \ac{llagn}.

Our findings suggest that the black hole shadow \citep{2000Falcke} of \ac{cena} would be visible in a bright, optically thin accretion flow at an observing frequency of a few THz. At this high frequency, a VLBI experiment above the Earth's troposphere would be able to resolve the $1.4 \pm 0.8 \,\mu\mathrm{as}$ shadow diameter with a minimal baseline length of $\sim 8000$\,km.

\section*{Acknowledgments}

\sloppy

The authors of the present paper thank the following
organizations and programs: the Academy
of Finland (projects 274477, 284495, 312496, 315721);
the Agencia Nacional de Investigaci\'on y Desarrollo (ANID),
Chile via NCN19\_058 (TITANs) and Fondecyt 3190878;
the Alexander von Humboldt Stiftung;
an Alfred P. Sloan Research Fellowship;
Allegro, the European ALMA Regional Centre node in the Netherlands, the NL astronomy
research network NOVA and the astronomy institutes of the University of Amsterdam, Leiden University and Radboud University;
the Black Hole Initiative at
Harvard University, through a grant (60477) from
the John Templeton Foundation; the China Scholarship
Council; Consejo
Nacional de Ciencia y Tecnología (CONACYT,
Mexico, projects  U0004-246083, U0004-259839, F0003-272050, M0037-279006, F0003-281692,
104497, 275201, 263356, 57265507);
the Delaney Family via the Delaney Family John A.
Wheeler Chair at Perimeter Institute; Dirección General
de Asuntos del Personal Académico-Universidad
Nacional Autónoma de México (DGAPA-UNAM,
projects IN112417 and IN112820);
the EACOA Fellowship of
the East Asia Core Observatories Association;
the European Research Council Synergy
Grant ``BlackHoleCam: Imaging the Event Horizon
of Black Holes" (grant 610058); the Generalitat
Valenciana postdoctoral grant APOSTD/2018/177 and
GenT Program (project CIDEGENT/2018/021); MICINN Research Project PID2019-108995GB-C22;
the
Gordon and Betty Moore Foundation (grants GBMF-
3561, GBMF-5278); the Istituto Nazionale di Fisica
Nucleare (INFN) sezione di Napoli, iniziative specifiche
TEONGRAV; the International Max Planck Research
School for Astronomy and Astrophysics at the
Universities of Bonn and Cologne; 
Joint Princeton/Flatiron and Joint Columbia/Flatiron Postdoctoral Fellowships, research at the Flatiron Institute is supported by the Simons Foundation; 
the Japanese Government (Monbukagakusho:
MEXT) Scholarship; the Japan Society for
the Promotion of Science (JSPS) Grant-in-Aid for JSPS
Research Fellowship (JP17J08829); the Key Research
Program of Frontier Sciences, Chinese Academy of
Sciences (CAS, grants QYZDJ-SSW-SLH057, QYZDJSSW-
SYS008, ZDBS-LY-SLH011). 

We further thank the Leverhulme Trust Early Career Research
Fellowship; the Max-Planck-Gesellschaft (MPG);
the Max Planck Partner Group of the MPG and the
CAS; the MEXT/JSPS KAKENHI (grants 18KK0090,
JP18K13594, JP18K03656, JP18H03721, 18K03709,
18H01245, JP19H01943, 25120007); the Malaysian Fundamental Research Grant Scheme (FRGS) FRGS/1/2019/STG02/UM/02/6; the MIT International Science
and Technology Initiatives (MISTI) Funds; the Ministry
of Science and Technology (MOST) of Taiwan (105-
2112-M-001-025-MY3, 106-2112-M-001-011, 106-2119-
M-001-027, 107-2119-M-001-017, 107-2119-M-001-020,
107-2119-M-110-005, 108-2112-M-001-048, and 109-2124-M-001-005);
the National Aeronautics and
Space Administration (NASA, Fermi Guest Investigator
grant 80NSSC20K1567, NASA Astrophysics Theory Program grant 80NSSC20K0527,
NASA grant NNX17AL82G, and Hubble Fellowship grant
HST-HF2-51431.001-A awarded by the Space Telescope
Science Institute, which is operated by the Association
of Universities for Research in Astronomy, Inc.,
for NASA, under contract NAS5-26555, and
NASA NuSTAR award 80NSSC20K0645); the National
Institute of Natural Sciences (NINS) of Japan; the National
Key Research and Development Program of China
(grant 2016YFA0400704, 2016YFA0400702); the National
Science Foundation (NSF, grants AST-0096454,
AST-0352953, AST-0521233, AST-0705062, AST-0905844, AST-0922984, AST-1126433, AST-1140030,
DGE-1144085, AST-1207704, AST-1207730, AST-1207752, MRI-1228509, OPP-1248097, AST-1310896,
AST-1337663, AST-1440254, AST-1555365, AST-1615796, AST-1715061, AST-1716327, AST-1716536,
OISE-1743747, AST-1816420, AST-1903847, AST-1935980, AST-2034306);
the Natural Science Foundation of China (grants 11573051, 11633006,
11650110427, 10625314, 11721303, 11725312, 11933007, 11991052, 11991053); a fellowship of China Postdoctoral Science Foundation (2020M671266); the Natural
Sciences and Engineering Research Council of
Canada (NSERC, including a Discovery Grant and
the NSERC Alexander Graham Bell Canada Graduate
Scholarships-Doctoral Program); the National Research
Foundation of Korea (the Global PhD Fellowship
Grant: grants 2014H1A2A1018695, NRF-2015H1A2A1033752, 2015-
R1D1A1A01056807, the Korea Research Fellowship Program:
NRF-2015H1D3A1066561, Basic Research Support Grant 2019R1F1A1059721); the Netherlands Organization
for Scientific Research (NWO) VICI award
(grant 639.043.513) and Spinoza Prize SPI 78-409;
the New Scientific Frontiers with Precision Radio Interferometry Fellowship awarded by the South African Radio Astronomy Observatory (SARAO), which is a facility of the National Research Foundation (NRF), an agency of the Department of Science and Innovation (DSI) of South Africa; the South African Research Chairs Initiative of the Department of Science and Innovation and National Research Foundation;
the Onsala Space Observatory
(OSO) national infrastructure, for the provisioning
of its facilities/observational support (OSO receives
funding through the Swedish Research Council under
grant 2017-00648) the Perimeter Institute for Theoretical
Physics (research at Perimeter Institute is supported
by the Government of Canada through the Department
of Innovation, Science and Economic Development
and by the Province of Ontario through the
Ministry of Research, Innovation and Science);
the Spanish Ministerio de Economía y Competitividad (grants
PGC2018-098915-B-C21, AYA2016-80889-P, PID2019-108995GB-C21); the State
Agency for Research of the Spanish MCIU through
the ``Center of Excellence Severo Ochoa'' award for
the Instituto de Astrofísica de Andalucía (SEV-2017-
0709); the Toray Science Foundation; the Consejería de Economía, Conocimiento, Empresas y Universidad of the Junta de Andalucía (grant P18-FR-1769), the Consejo Superior de Investigaciones Científicas (grant 2019AEP112);
the US Department
of Energy (USDOE) through the Los Alamos National
Laboratory (operated by Triad National Security,
LLC, for the National Nuclear Security Administration
of the USDOE (Contract 89233218CNA000001);
the European Union’s Horizon 2020
research and innovation programme under grant agreement
No 730562 RadioNet; ALMA North America Development
Fund; the Academia Sinica; Chandra TM6-
17006X and DD7-18089X; the GenT Program (Generalitat Valenciana)
Project CIDEGENT/2018/021. 

This work used the
Extreme Science and Engineering Discovery Environment
(XSEDE), supported by NSF grant ACI-1548562,
and CyVerse, supported by NSF grants DBI-0735191,
DBI-1265383, and DBI-1743442. XSEDE Stampede2 resource
at TACC was allocated through TG-AST170024
and TG-AST080026N. XSEDE JetStream resource at
PTI and TACC was allocated through AST170028.
The simulations were performed in part on the SuperMUC
cluster at the LRZ in Garching, on the
LOEWE cluster in CSC in Frankfurt, and on the
HazelHen cluster at the HLRS in Stuttgart. This
research was enabled in part by support provided
by Compute Ontario (http://computeontario.ca), Calcul
Quebec (http://www.calculquebec.ca) and Compute
Canada (http://www.computecanada.ca). We thank
the staff at the participating observatories, correlation
centers, and institutions for their enthusiastic support.

This paper makes use of the following ALMA data: ADS/JAO.ALMA\#2016.1.01198.V.
ALMA is a partnership
of the European Southern Observatory (ESO;
Europe, representing its member states), NSF, and
National Institutes of Natural Sciences of Japan, together
with National Research Council (Canada), Ministry
of Science and Technology (MOST; Taiwan),
Academia Sinica Institute of Astronomy and Astrophysics
(ASIAA; Taiwan), and Korea Astronomy and
Space Science Institute (KASI; Republic of Korea), in
cooperation with the Republic of Chile. The Joint
ALMA Observatory is operated by ESO, Associated
Universities, Inc. (AUI)/NRAO, and the National Astronomical
Observatory of Japan (NAOJ). The NRAO
is a facility of the NSF operated under cooperative agreement
by AUI. APEX is a collaboration between the
Max-Planck-Institut f{\"u}r Radioastronomie (Germany),
ESO, and the Onsala Space Observatory (Sweden). The
SMA is a joint project between the SAO and ASIAA
and is funded by the Smithsonian Institution and the
Academia Sinica. The JCMT is operated by the East
Asian Observatory on behalf of the NAOJ, ASIAA, and
KASI, as well as the Ministry of Finance of China, Chinese
Academy of Sciences, and the National Key R\&D
Program (No. 2017YFA0402700) of China. Additional
funding support for the JCMT is provided by the Science
and Technologies Facility Council (UK) and participating
universities in the UK and Canada.
The
LMT is a project operated by the Instituto Nacional
de Astrófisica, Óptica, y Electrónica (Mexico) and the
University of Massachusetts at Amherst (USA), with financial support from the Consejo Nacional de Ciencia y Tecnología and the National Science Foundation.
The IRAM 30-m telescope on Pico Veleta, Spain is operated
by IRAM and supported by CNRS (Centre National de
la Recherche Scientifique, France), MPG (Max-Planck-
Gesellschaft, Germany) and IGN (Instituto Geográfico
Nacional, Spain). The SMT is operated by the Arizona
Radio Observatory, a part of the Steward Observatory
of the University of Arizona, with financial support of
operations from the State of Arizona and financial support
for instrumentation development from the NSF.
The SPT is supported by the National Science Foundation
through grant PLR- 1248097. Partial support is
also provided by the NSF Physics Frontier Center grant
PHY-1125897 to the Kavli Institute of Cosmological
Physics at the University of Chicago, the Kavli Foundation
and the Gordon and Betty Moore Foundation grant
GBMF 947. The SPT hydrogen maser was provided on
loan from the GLT, courtesy of ASIAA. The EHTC has
received generous donations of FPGA chips from Xilinx
Inc., under the Xilinx University Program. The EHTC
has benefited from technology shared under open-source
license by the Collaboration for Astronomy Signal Processing
and Electronics Research (CASPER). The EHT
project is grateful to T4Science and Microsemi for their
assistance with Hydrogen Masers. This research has
made use of NASA’s Astrophysics Data System. We
gratefully acknowledge the support provided by the extended
staff of the ALMA, both from the inception of
the ALMA Phasing Project through the observational
campaigns of 2017 and 2018. We would like to thank
A. Deller and W. Brisken for EHT-specific support with
the use of DiFX. We acknowledge the significance that
Maunakea, where the SMA and JCMT EHT stations
are located, has for the indigenous Hawaiian people.

The grants listed above collectively fund the Event Horizon Telescope project.

\section*{Author contributions}

K.A., L.B., C.-k.C., S.I., M.J., J.K., J.Y.K., T.P.K., J.L., E.L., D.P.M, V.R., K.L.J.R., I.V.B, and M.W. have worked on the calibration of the \ac{eht} data.
K.A., K.L.B., A.C., J.L.G., S.I., M.J., M.D.J., C.N., D.W.P., F.R., and M.W. have worked on the image reconstruction.
M.B., K.C., J.D., P.G.E., H.F, C.M.F, C.G., M.J., M.K., Y.M., A.M., S.M, E.R., and M.W. have worked on the interpretation of the results.
M.J, M.K., C.M., and E.R. have coordinated the research.
The Event Horizon Telescope collaboration as a whole has enabled this research by building the EHT instrument and producing the tools and knowledge for the reduction, analysis, and interpretation of the data.

\section*{Competing interests}

The authors declare no competing interests.

\section*{Main text figures}

\begin{figure}[h!]
\centering
\includegraphics[width=0.85\textwidth]{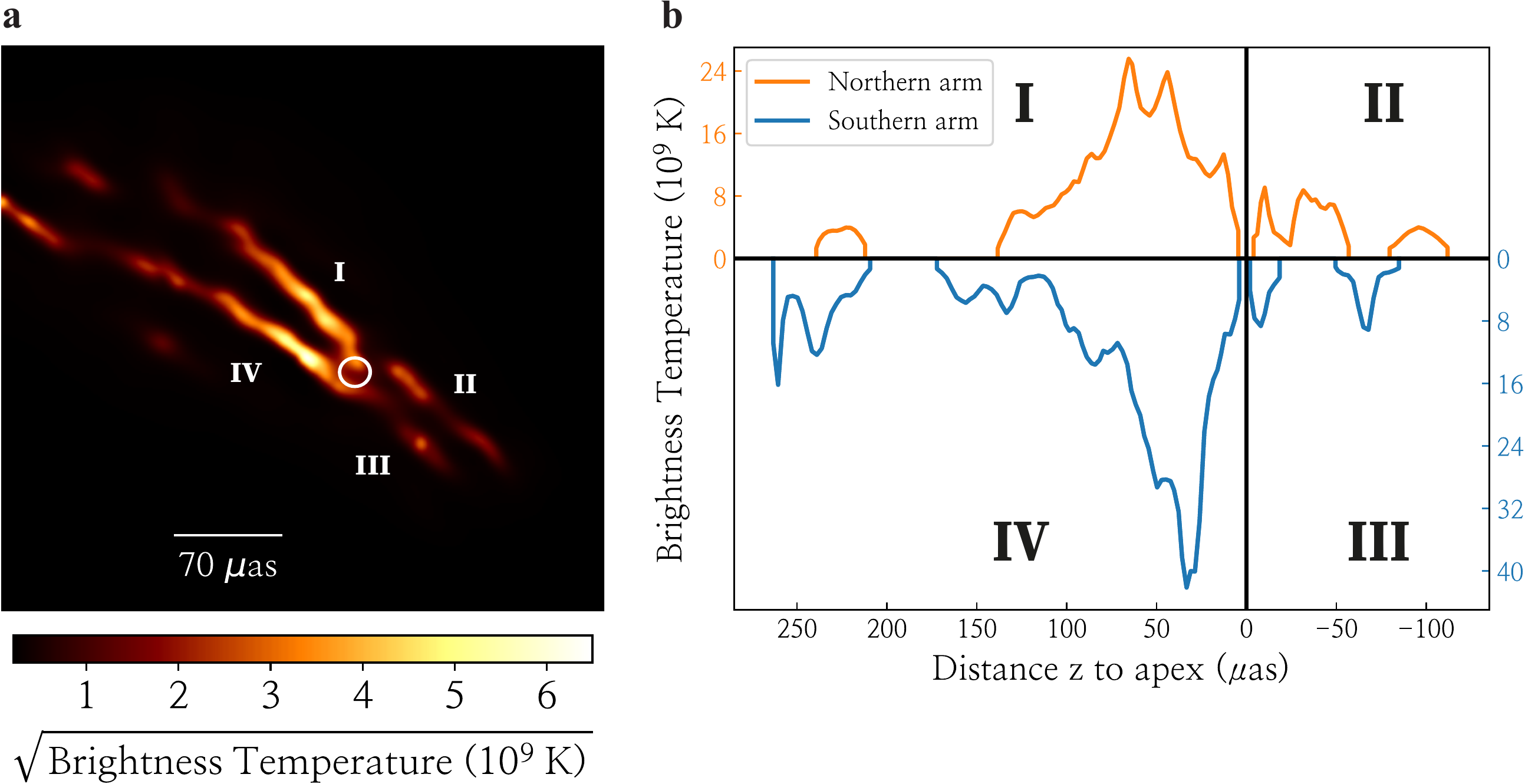} 
\caption{\textbf{Image analysis of the final model.}
                 The model image shown in the left panel (a) corresponds to the image from \autoref{CenA-fig:data-model} with a pixel size of 2\,\ac{muas}.
                 Here, the tentative position of the jet apex is indicated with a circle. The size of the circle indicates the uncertainty in the apex location.
                 The right panel (b) shows the central brightness temperatures along the jet ridgelines from the model in four quadrants as a function of distance to the jet apex.
                 The quadrants I, II, III, and IV correspond to the similarly marked regions of the jet in the left panel.
                 Negative values for the distance to the jet apex are assigned for the counterjet region. Brightness temperatures of the fainter NW (orange line) and brighter SE (blue line) arms are shown in the upper and lower panels, respectively.
                 }
\label{CenA-fig:im-analysis}
\end{figure}

\begin{figure}[h!]
\centering
\includegraphics[width=0.7\textwidth]{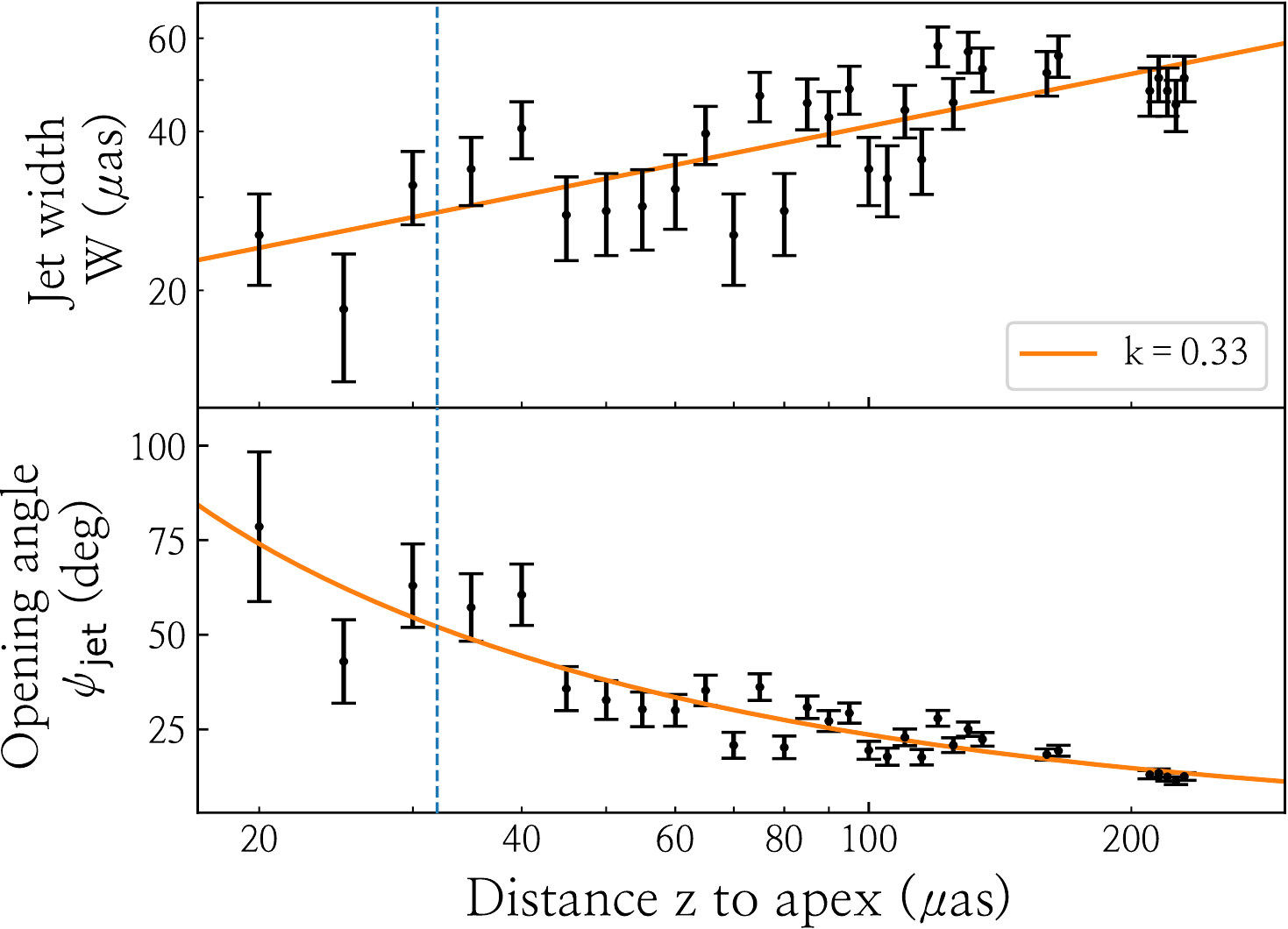}
\caption{\textbf{\ac{cena} collimation profile derived from the final model.}
                 The top panel shows the jet width $W$ as a function of apex distance $z$, overplotted with a $W(z) = A z^k$ least-squares fit, with $k=0.33 \pm 0.05{\mid}_{\mathrm{stat}} \pm 0.06{\mid}_{\mathrm{sys}}$ (Methods~\ref{sec:methods:col}). The systematic error has been derived based on the uncertainty in the apex position. The error bars are derived from a 5\,\ac{muas} one sigma statistical uncertainty on each pixel in the model.
                 The bottom panel shows the corresponding jet opening angle between the apex and the two jet arms, which becomes unreliable at $z \lesssim z_\mathrm{col} = 32\,\ac{muas}$ due to instrumental resolution limitations (see text). The transition region is indicated with a vertical blue dashed line. 
                 A base-10 logarithmic scale is used for the horizontal axis.
                 }
\label{CenA-fig:jet_fit}
\end{figure}

\section{Methods}

\subsection{Processing of observational data}
\label{sec:obsandcal}

This section describes the 2017 \ac{eht} observations of \ac{cena}, the model-independent calibration \citep{Janssen2019} with two separate pipelines, the flux density calibration, and known measurement issues and systematics with corresponding mitigation strategies.
The final data sets coming out from the two pipelines are both used for the scientific analysis as cross-verification (Supplementary Fig.~\ref{CenA-SuppFIG:image-consistency}).

\subsubsection{Data acquisition}
\begin{figure}[h]
\centering
\includegraphics[width=0.8\textwidth]{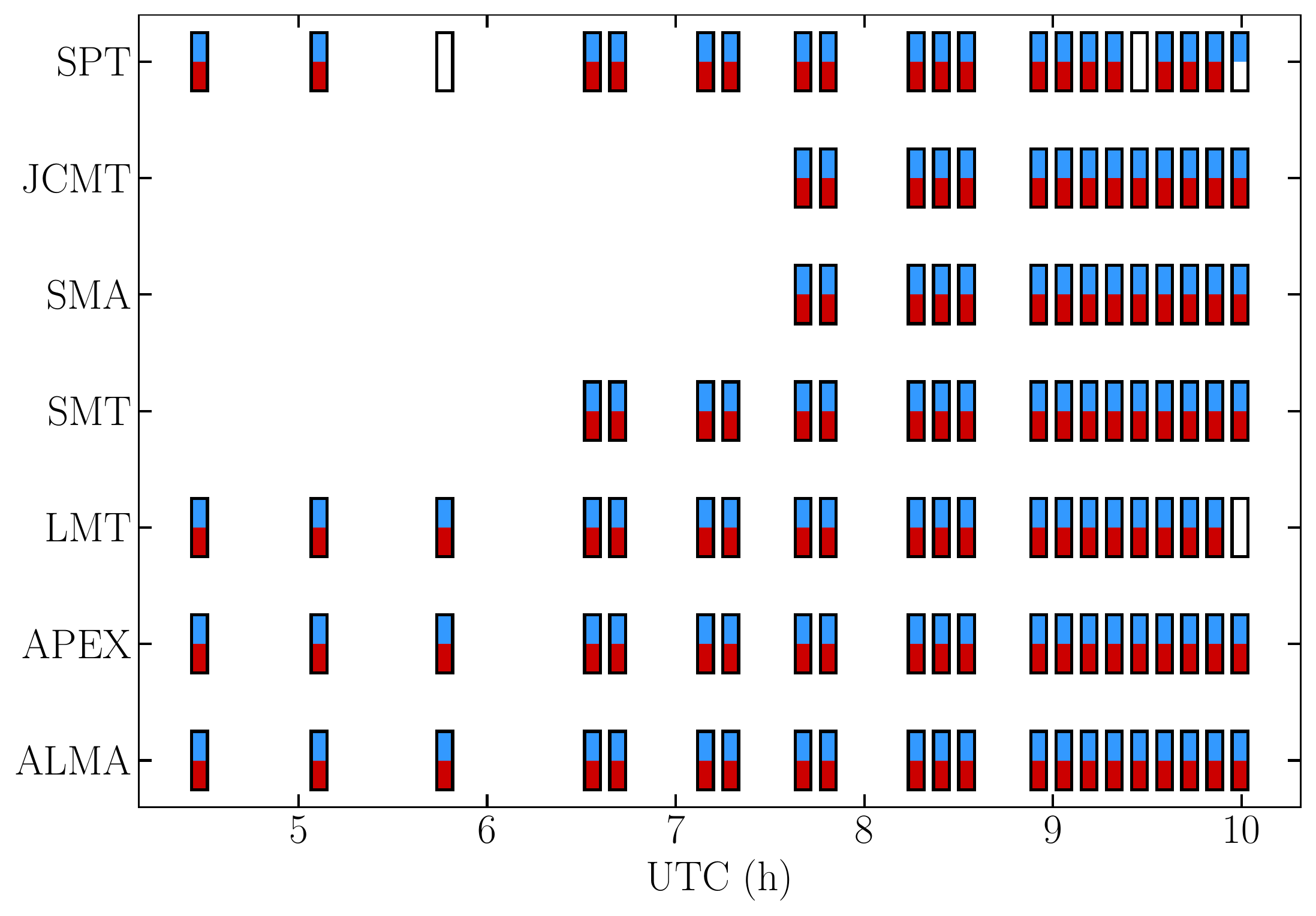}
\caption{\textbf{\ac{cena} observing block.}
                 The boxes show the planned VLBI scans in the observational VEX file.
                 Blue (top) and red (bottom) markers in each box show detections in the EHT-HOPS and rPICARD data, respectively.}
\label{CenA-SuppFIG:sched}
\end{figure}

\ac{cena} (PKS\,1322$-$428, hosted in the NGC\,5128 elliptical galaxy, $\alpha_\mathrm{J2000} = 13^\mathrm{h}25^\mathrm{m}27.62^\mathrm{s}$, $\delta_\mathrm{J2000} = \ang{-43;01;08.81}$) was observed by the \ac{eht} in a six hour long track on April 10 2017, with a total on-source integration time of 105\,minutes (Supplementary~Fig.~\ref{CenA-SuppFIG:sched}). The observations were carried out by the \ac{aa}, \ac{ap}, \ac{jc}, \ac{lm}, \ac{sp} \citep{2018KimSPT}, \ac{sm}, and \ac{az} \citep{eht-paperII}. For ALMA, 37 of the 12\,m dishes were phased-up \citep{Matthews2018}. The \ac{eht} IRAM~30\,m Telescope is not able to see \ac{cena} jointly with rest of the array due to the low declination of the source. The data were recorded on Mark\,6 \ac{vlbi} recorders \citep{Whitney2013} with 2-bit sampling in two 2\,GHz wide bands, `low' and `high', centred around 227.1\,GHz and 229.1\,GHz, respectively.
Unless stated otherwise, results are derived using the combined low+high band data.
Quarter-wave plates at each site except \ac{aa} were used to observe circularly polarized light. The data were correlated with the \difx{} software \citep{Deller2007,Deller2011}. The \polconvert{} \citep{MartiVidal2016} software was used to convert the phased \ac{aa} data \citep{Matthews2018} from a linear polarization basis to a circular basis after correlation, based on solutions from the calibration of the connected-element \ac{aa} data \citep{Goddi2019QA2}.

\subsubsection{Data reduction pipelines}
\label{sec:methods:pipelines}
The autocorrelation normalization, feed angle rotation, fringe fitting, bandpass calibration, and a priori correction of atmospheric phase turbulence \citep{2017Thompson} were performed independently by the CASA-based rPICARD \citep{McMullin2007,Janssen2019} and EHT-HOPS \citep{Whitney_2004,Blackburn2019} pipelines.
\difx{} produces FITS-IDI and Mark4 data. rPICARD uses the FITS-IDI product and converts it into the Measurement Set format. EHT-HOPS uses the Mark4 data.
Both software packages convert the calibrated data into the UVFITS format for further processing.

rPICARD performs an upstream correction for the feed rotation angle and uses station-based global fringe fitting based on an unpolarized point source model to correct for phases, delays, and rates consistently for the right circular polarization and left circular polarization signal paths \citep{Schwab1983}.
Atmospheric phase and residual delay variations are corrected within the expected coherence time by fringe fitting segmented data of each \ac{vlbi} scan.
The segmentation length is set by the \ac{sn} of each baseline.

For EHT-HOPS, the feed rotation angle is corrected after the fringe-fitting together with an additional polarization calibration step, where complex R-L gain offsets are solved for.
Delays and rates are found in a baseline-based fringe search and referenced to individual stations with a least-squares optimization \citep{alef1986}.
Atmospheric phases are corrected by fitting a polynomial phase model to the data on baselines to the most sensitive reference station in each scan.
A round-robin approach is used to avoid fitting to thermal noise and the degree of the polynomial is set by the \ac{sn} of the data.

\subsubsection{Gain amplitude calibration}
The flux density calibration is done based on determined station sensitivities in a common framework for the rPICARD and EHT-HOPS data \citep{eht-paperIII}.
The sensitivity of a station $i$ is given by its system equivalent flux density ($\mathrm{SEFD}_i$) in \ac{jy}, which takes into account the gain and total noise power along a telescope's signal chain as a function of time $t$ and frequency $\nu$. On a baseline $i$--$j$, correlation coefficients $\xi_{(i,j)}$ in units of thermal noise are calibrated to a physical radiation intensity scale of correlated flux density $\mathcal{S}_{(i,j)}$ through
\begin{equation}
\mathcal{S}_{i,j}(t, \nu) = \frac{\sqrt{\mathrm{SEFD}_i(t, \nu) \mathrm{SEFD}_j(t, \nu)}}{\eta_Q} \, \xi_{i,j}(t, \nu) \, ,
\end{equation}
where $\eta_Q$ is the quantization efficiency. For data recorded with 2-bit sampling, we have $\eta_Q \approx 0.88$.

The gains of co-located stations were solved based on a contemporaneous measurement of the total flux density $S_0 = 5.62$\,\ac{jy} of the source with the \ac{aa} interferometer \citep{Goddi2019QA2,Blackburn2019,eht-paperIII}.
The correlated flux $\mathcal{S}$ measured between two co-located sites $p$ and $q$ should be equal to $S_0$ and for a third station $o$, we should have $\mathcal{S}_{po} = \mathcal{S}_{qo}$. It follows, that we can solve for the station-based amplitude gains $\mathcal{A}$ of $p$ and $q$ with a self-calibration approach. Here, the model is given by the constant flux density $S_0$ seen by baselines between co-located sites. No gain corrections for non-co-located (`isolated') stations are solved for. 

Ad hoc correction factors are used to correct signal losses at \ac{ap} due to an injected instrumental signal and at \ac{sm} due to temporary losses of bandwidth \citep{eht-paperIII}.
Additionally, \ac{lm} and \ac{sp} suffered from pointing problems, which result in significant amplitude variations between and within \ac{vlbi} scans.
These losses cannot be estimated a priori and must be corrected with self-calibration gain solutions derived within short $\sim 10$\,s segments from high \ac{sn} data.
The \ac{az} station was able to track the source down to an elevation of a few degrees.
Large self-calibration gain factors are therefore needed towards the end of the experiment.
Besides these known data issues, gain corrections factors are well constrained within a determined a priori error budget ranging between 10\,--\,20\,\% for the individual stations \citep{eht-paperIII}.

\subsection{Imaging}
\label{sec:imaging}

In this section, we describe how we obtained our image model from the observational data.
In a first step, we have established a blind consensus between different imaging methods.
Then, we have fine-tuned the parameters of one method, eht-imaging \citep{Chael_2016,Chael_2018_Imaging}, for the rPICARD and EHT-HOPS data to obtain final images for the analysis of the \ac{cena} jet structure.

The highest resolution images of this southern source prior to this work were obtained within the TANAMI program \citep{Ojha2010} at 8 and 22\,GHz with a maximum resolution of 400\,\ac{muas}, showing an extremely collimated structure with multiple distinct radio knot emission regions \citep{Mueller2014}.
In a previous single-baseline non-imaging study of \ac{cena}, a bright compact core was detected at 215\,GHz \citep{2018Kim}.

\subsubsection{Blind challenge}
Similarly to the method used when the shadow of M\,87* was resolved by the \ac{eht} \citep{eht-paperI,eht-paperIV}, we have carried out a \textit{blind imaging challenge} before proceeding to the scientific analysis of the data.
In this challenge, a number of individuals have reconstructed an image of the source independently of each other.
Early (not fully verified) low band data from the EHT-HOPS pipeline was used, which had slightly larger amplitude gain errors from outdated a priori calibration parameters.
Out of twelve total images, six had acceptable reduced $\chi^2_\mathrm{cp} < 2$ for the closure phases.
These images were obtained with the eht-imaging and SMILI \citep{Akiyama_2017a,Akiyama_2017b} regularized maximum likelihood methods and the Difmap \citep{Shepherd1994,Shepherd_1997} and CASA \citep{Rau2011,Janssen2019} CLEAN methods \citep{1974Hoegbom,1980Clark}.
The images that did not make the $\chi^2$ cut often showed spurious emission features and strong sidelobe structures.

\subsubsection{Final imaging method}
\begin{figure}[h!]
\centering
\includegraphics[width=\textwidth]{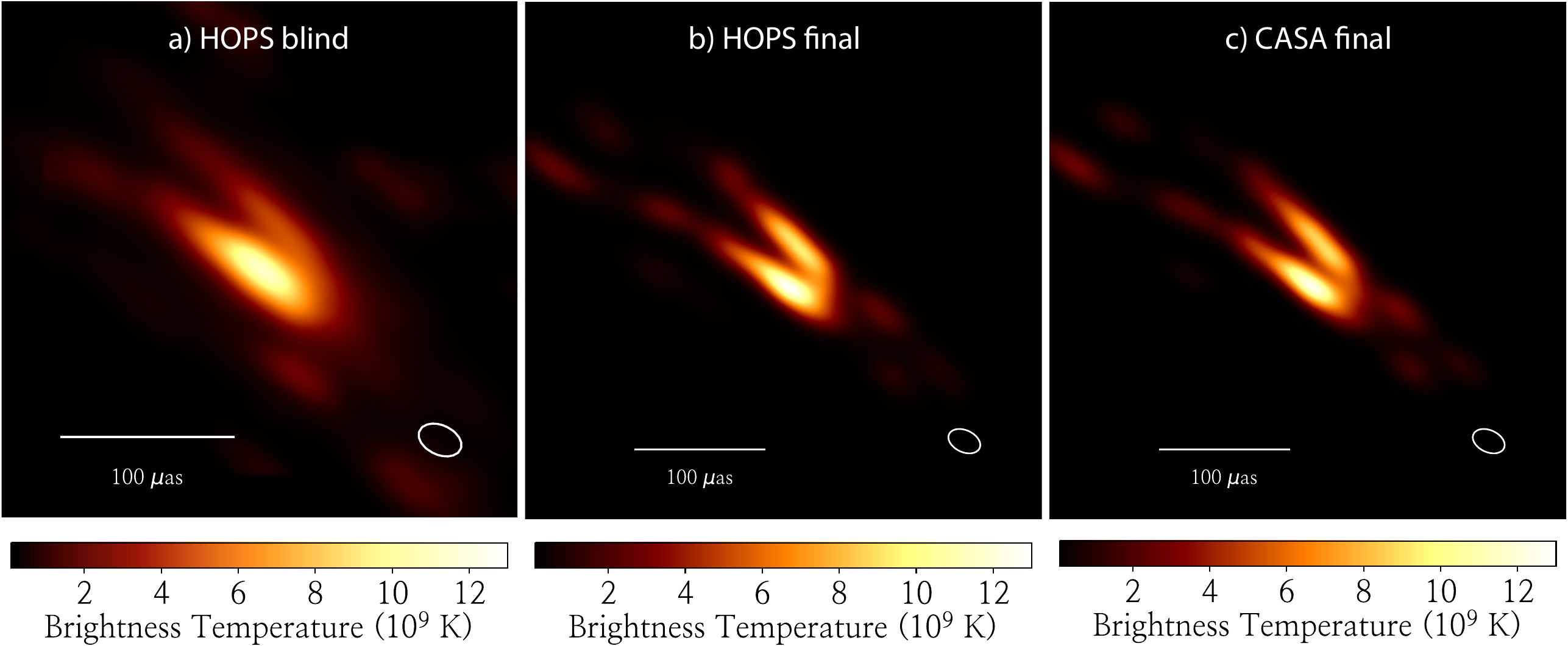}
\caption{\textbf{Image consistency across data sets.}
                 Left (a): Average image of six blind reconstructions from early EHT-HOPS low band (226\,--228\,GHz) data, which was prepared with outdated a priori calibration parameters.
                 Middle (b) and right (c): Images from the EHT-HOPS and rPICARD data, respectively, reconstructed with the final imaging parameters (Table~1).
                 The restoring beams are plotted in the bottom right corner.}
\label{CenA-SuppFIG:image-consistency}
\end{figure}

With the imaging challenge, we have established that different methods converge towards the same robust source structure (Supplementary~Fig.~\ref{CenA-SuppFIG:image-consistency}), independent of shared human bias.
Further imaging analysis of the rPICARD \& EHT-HOPS science release data was pursued with the final M\,87* eht-imaging script \citep{eht-paperIV}, which is based on application of a regularized maximum likelihood method that includes a maximum-entropy term.
Using a second-moment-based pre-calibration, \ac{lm} gains were stabilized with respect to the better constrained \ac{az} amplitudes \citep{2019Issaounsecondmoment}. As \ac{cena} is sufficiently compact within the \ac{eht} beam, the short \ac{lm}\,--\,\ac{az} baseline measures a Gaussian-like source structure. 
We have performed an initial self-calibration to a Gaussian with size $\Theta_\mathrm{maj} \times \Theta_\mathrm{min}$ at a position angle $\Theta_\mathrm{PA}$ and with a total flux of $S_G$.
Any gains that were erroneously introduced in this process can later be reconciled in image-based self-calibration steps.
To solve for the image brightness distribution $\mathcal{I}$ with a regularized maximum likelihood method (employed by eht-imaging), we are minimizing
\begin{equation}
\sum_\mathrm{D} \alpha_D \chi^2_D(\mathcal{I}) \;\;-\;\; \sum_\mathrm{R} \beta_R \Lambda_R(\mathcal{I})\,.
\end{equation}
Here, $D$ represents the collection of data terms, which are derived from the measured visibilities and have approximately normal noise statistics \citep{2020Blackburn}: amplitudes, closure phases, and log closure amplitudes.
Corresponding to each data term, we have a goodness-of-fit function $\chi^2_D = \{\chi^2_\mathrm{amp}, \chi^2_\mathrm{cp}, \chi^2_\mathrm{lca}\}$ and relative weighting $\alpha_D = \{\alpha_\mathrm{amp}, \alpha_\mathrm{cp}, \alpha_\mathrm{lca}\}$,
We have performed four incremental imaging runs with subsequent self-calibration, over which we have increased the weight of each data term: $\alpha_D^{(1)} \rightarrow \alpha_D^{(2)} \rightarrow  \alpha_D^{(3)}$ with $\alpha_D^{(1)} < \alpha_D^{(2)} < \alpha_D^{(3)}$ $\forall D$.
Regularizer terms $\Lambda_R$ are included with weights $\beta_R$ to impose additional assumptions on the image.
We have imposed two regularization parameters:
One for a maximum entropy method (MEM) \citep{1986Narayan} term with weight $\beta_\mathrm{MEM}$ and another one for the amount of compact flux $Z_0$ in the image with weight $\beta_z$.
The MEM term minimizes the entropy of $\mathcal{I}$ with respect to a prior image $\Phi$, which results in a similarity between the two images for each pixel $i$. Here, we used $\Lambda_\mathrm{MEM} = - S_0^{-1} \sum_\mathrm{i} \mathcal{I}_i \log{(\mathcal{I}_i/\Phi_i)}$.
For the MEM prior image $\Phi$, we have chosen a Gaussian model oriented along the direction of the large-scale jet, which we also used as initialization for our imaging.
It is expected that $Z_0 < S_0$, as a significant portion of the flux measured by ALMA may come from different emission mechanisms and larger scales outside of the EHT field of view. 
In fact, the $\sim$\,150\,m \ac{jc}-\ac{sm} baseline sees a flux density of about 5\,Jy and at 2\,km, \ac{aa}-\ac{ap} recovers only $\sim$\,4\,Jy.
For M\,87*, the EHT measured $Z_0 \approx S_0 / 2$ \citep{eht-paperIV}.

The numerical values of the final imaging parameters are given in the supplementary material (Table~1). Optimal parameters were chosen based on an empirical minimization of $\chi^2_\mathrm{D}$, median station gains $\mathcal{A}^\mathrm{(sc)}$ from self-calibration, and patches of spurious flux in the image.
Additionally, we took the similarity of image reconstructions from the rPICARD and EHT-HOPS data for a given set of parameters into account to avoid over-fitting to data peculiarities that result from assumptions made during the data calibration.
A variety of images that can be reconstructed with various combinations of the free imaging parameters can be shared upon reasonable requests.
We have chosen for an eht-imaging reconstruction of the rPICARD data for our final image, as this imaging method and data set have been studied most extensively.

Our images are shown in units of brightness temperature $\mathcal{T}$\,[K], which is related to a flux density $S$ in \ac{jy} through the observing wavelength $\lambda$, Boltzmann constant $k_\mathrm{B}$, and angular resolution element $\Omega$ as $\mathcal{T} = \lambda^2 \left(2 k_\mathrm{B} \Omega\right)^{-1} S$.

Fundamental data properties and fits of the final image model to the data are shown in Supplementary~Fig.~\ref{CenA-fig:dataplt}. In Supplementary~Fig.~\ref{CenA-SuppFIG:projplot}, we show the measured amplitudes projected along and perpendicular to the jet \ac{pa}.
Along the jet axis, amplitudes fall off quickly at long projected baseline lengths, indicating the absence of substructures along the jet.
Perpendicular to the jet, `bouncing' amplitudes out to large projected baseline lengths occur, due to the strong intensity gradients across the transverse jet profile.

\subsection{Synthetic data imaging tests}
\label{sec:symba-tests}

We have used the SYMBA software \citep{2020Roelofs} to perform imaging studies based on simulated observations.
Given an input source model $\mathcal{M}$, SYMBA follows the entire \ac{eht} signal path to predict which source structure $\mathcal{I}$ would eventually be reconstructed.
Thereby, we can assess how close our image reconstruction comes to the ground truth structure $\mathcal{M}$ of a fabricated observed source.
SYMBA simulates the parameters of the Earth's atmosphere with the ATM module \citep{Pardo2001} to add sky noise, signal attenuation, and phase turbulence.
Next, gain, leakage, pointing, and focus errors plus thermal noise are introduced for each telescope in the array based on known telescope properties \citep{Blecher2017}.
Afterwards, the simulated corrupted data are calibrated by rPICARD in the same way as observational data.
The $u,v$ coverage and SEFD sensitivities are taken from the 2017 \ac{cena} \ac{eht} observation track.
The simulated calibrated data are then imaged with the same final eht-imaging script used to image the observational data in this work.

\begin{figure}[h!]
\centering
\includegraphics[width=0.33\textwidth]{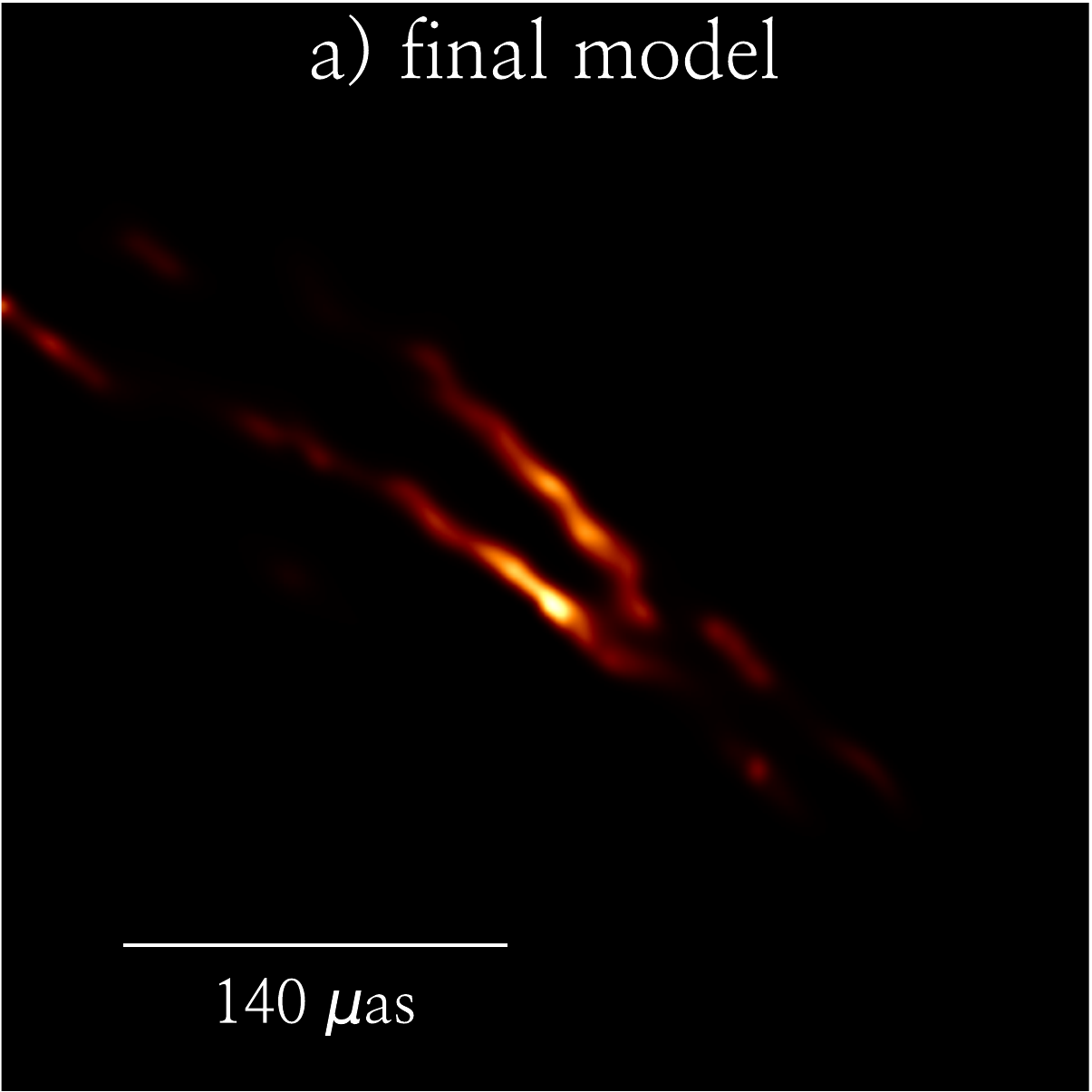}
\includegraphics[width=0.33\textwidth]{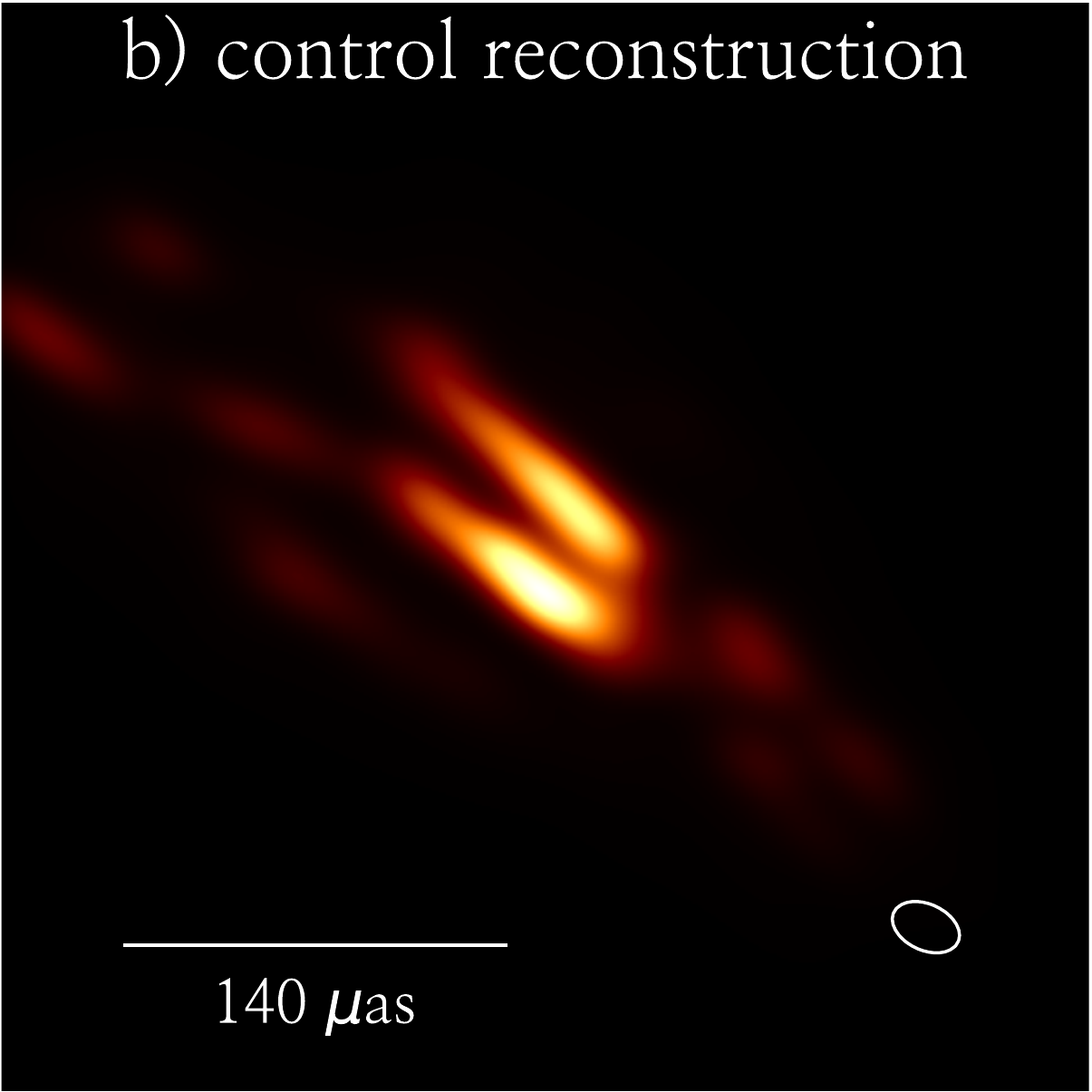} \\
\includegraphics[width=0.33\textwidth]{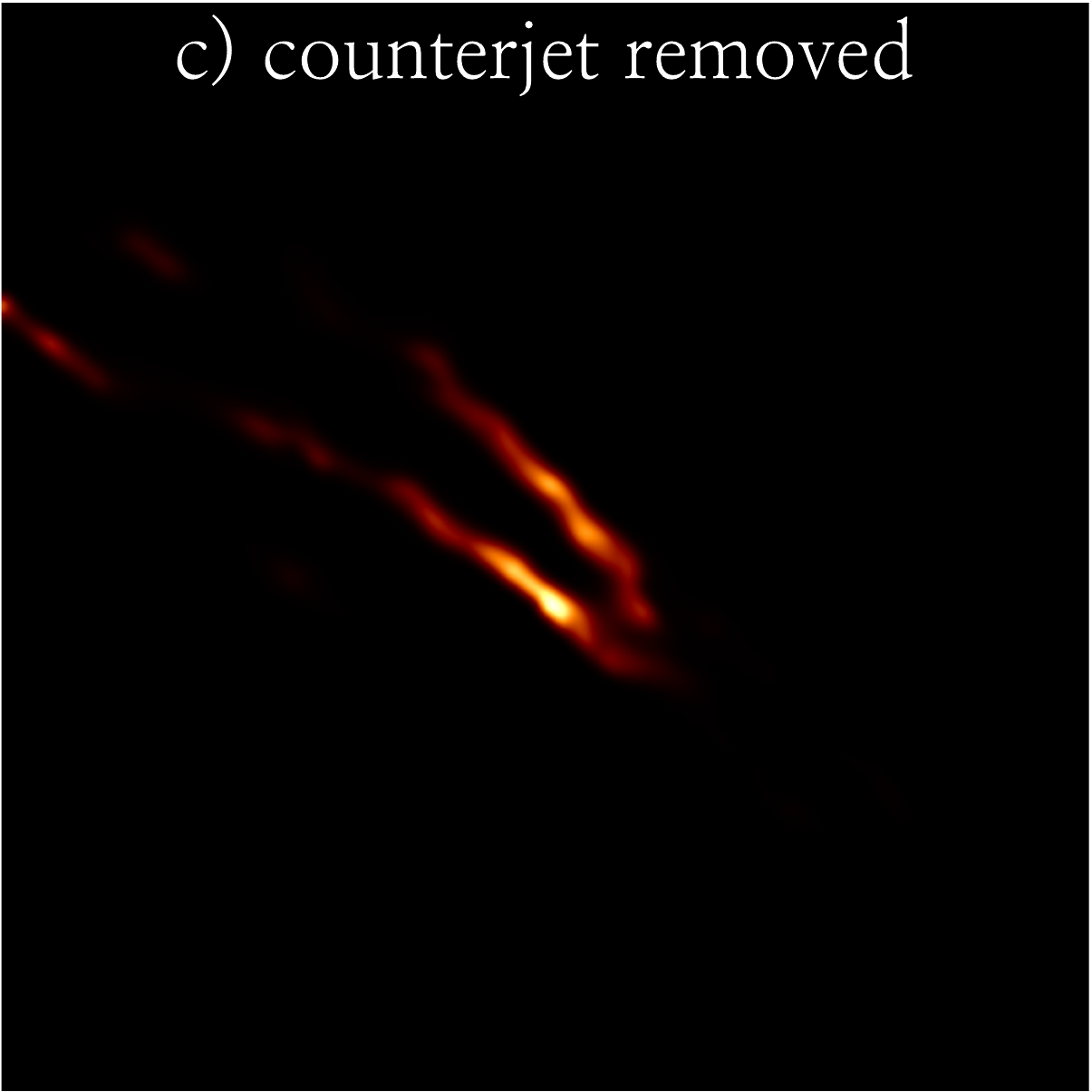}
\includegraphics[width=0.33\textwidth]{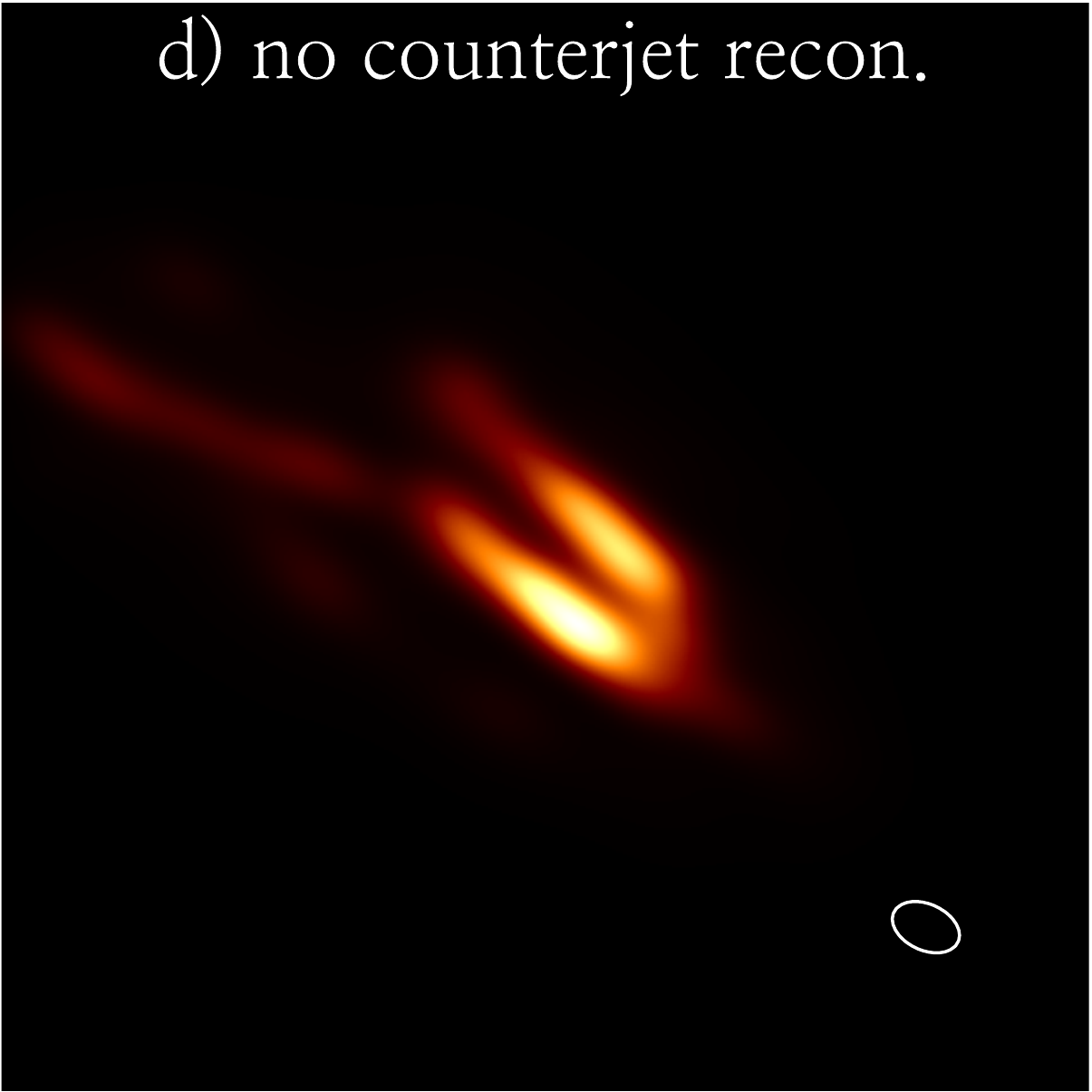} \\
\includegraphics[width=0.33\textwidth]{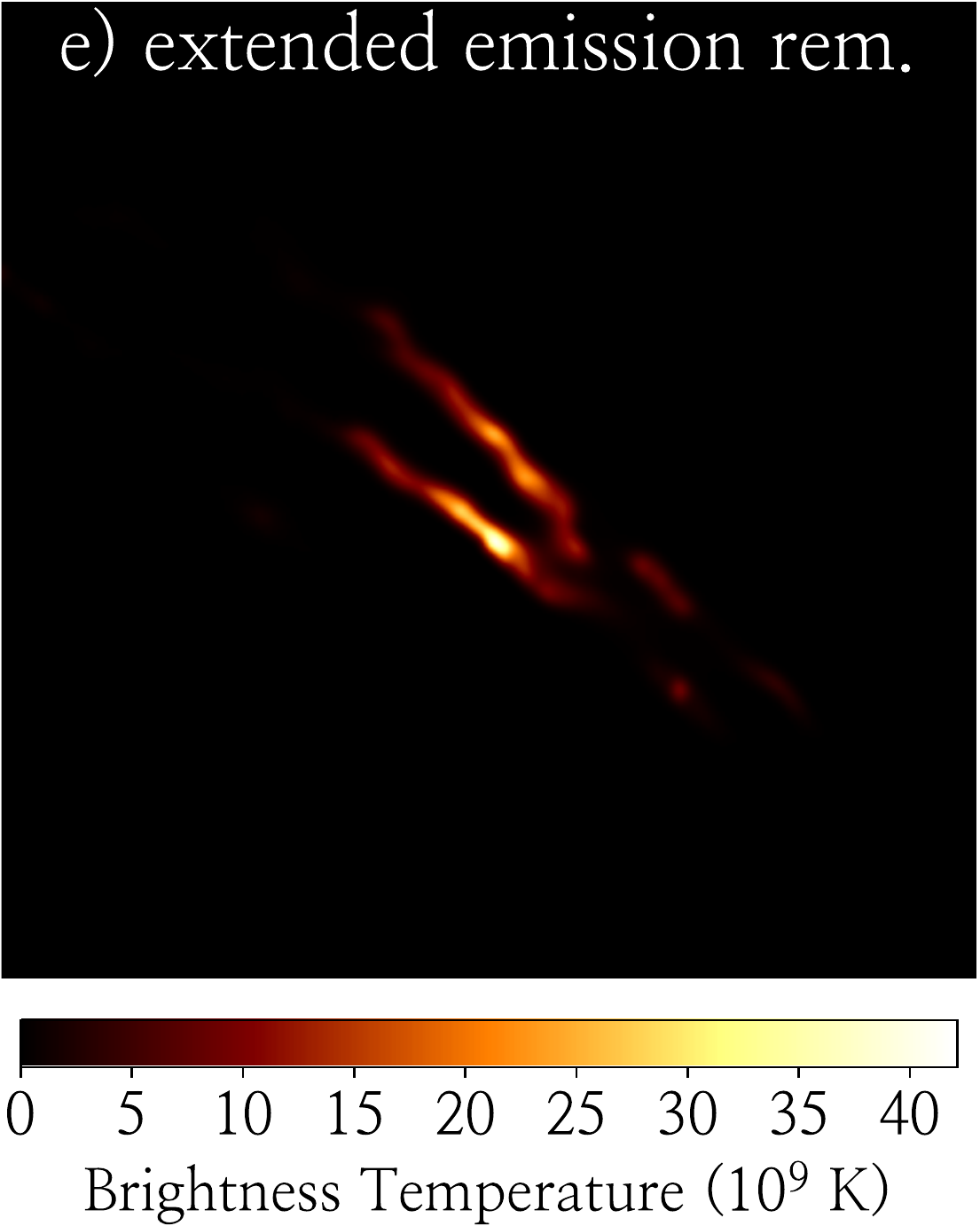}
\includegraphics[width=0.33\textwidth]{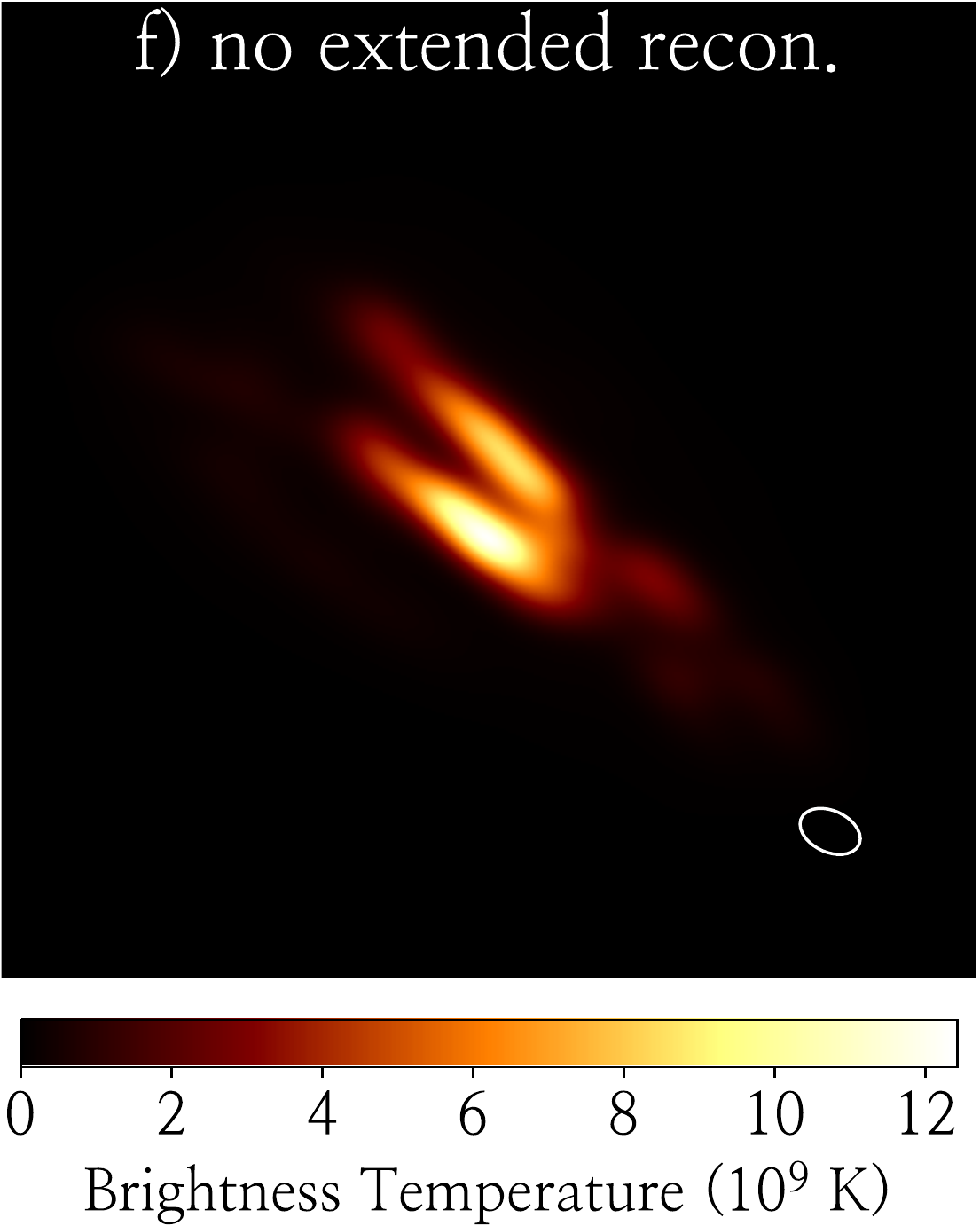} 
\caption{\textbf{SYMBA synthetic data imaging tests.}\
         The left panel shows the input models $\mathcal{M}$, which are based on our observational image reconstruction. The corresponding reconstructions $\mathcal{I}$ from the SYMBA pipeline are displayed in the right panel.
         (a-b) a control study, where the final model $\mathcal{M}_\mathrm{final}$ underlying the image reconstruction $\mathcal{I}_\mathrm{final}^\mathrm{(obs)}$ from the observational rPICARD data is passed through SYMBA.
         (c-d and e-f) same as (a-b) but the counterjet and extended jet emission features have been removed.
         The SYMBA reconstructions are convolved with a restoring beam, which matches the nominal resolution of the observation (shown in the bottom right corner).
        }
\label{CenA-SuppFIG:symba}
\end{figure}

To assess the robustness of secondary features in our image reconstruction, we have performed three synthetic data tests (Supplementary~Fig.~\ref{CenA-SuppFIG:symba}).
Firstly, a control study to demonstrate that the output reconstruction from SYMBA correctly matches the input model $\mathcal{M}_\mathrm{final}$.
Then, we have removed the counterjet and emission features at large distance to the apex $z$ from $\mathcal{M}_\mathrm{final}$ to verify that these do not spuriously appear in our simulated observation.

Furthermore, we have explored the upper limit on the brightness $\mathcal{T}_\mathrm{sp}$ of potential emission from the jet spine by adding a weak emission component in the central jet region to $\mathcal{M}_\mathrm{final}$.
The goal was to find the smallest $\mathcal{T}_\mathrm{sp}$, which would still be registered as an emission region in the reconstructed image $\mathcal{I}\left(\mathcal{M}_\mathrm{final}+\mathcal{T}_\mathrm{sp}\right)$.

\subsection{Jet structure analysis}
\label{sec:jetstructure}

This section describes how we extract fundamental jet parameters from our image based on geometric arguments.

\subsubsection{The position of the jet apex}
\label{sec:methods:apexpos}

We can empirically determine the approximate position of the jet apex, where the jet and counterjet are being launched, from the high resolution image model shown in \autoref{CenA-fig:im-analysis}.
A zoomed-in version of this plot is shown in Supplementary~Fig.~\ref{CenA-SuppFIG:apex-loc}, which is overlaid with visual aids for the determination of the apex location.
The first consideration is that the apex should be located in the region where the streamlines of the approaching jet converge. In our image, this convergence region lies upstream of the optically thick radio cores (assumed to correspond to the brightest regions of the jet) for both the NW and SE jet arms. Here, we are limited by the resolution of our instrument, but a tentative merge of the two arms can be seen.
The upper arm (region \texttt{I} in the figure) exhibits a strong bend, while the lower arm (\texttt{IV}) remains mostly straight. We note that a similar structure, where one jet arm appears to be straighter than the other one is also present in the M\,87 jet \citep{Walker2018}.
The second consideration is the symmetry between the approaching jet and the counterjet. We note that there is no clear correspondence between individual features in the jet and counterjet. The counterjet appears straight with two components in the upper region (\texttt{II}) and one component in the lower region (\texttt{III}). As the apex must be upstream of the counterjet, the closest component of the receding jet to the approaching jet constrains how far upstream of the approaching jet the apex position can be.
In fact, the position we assume for the apex based on the first consideration, where the streamlines of the approaching jet converge, lies halfway between the radio core in region \texttt{I} and the closest counterjet component in region \texttt{II}.
It should be noted that a simple extrapolation of only the edge-brightened approaching jet would place the apex well inside the faint counterjet region.

Based on the robustness of our image reconstructions with different data sets, software packages, and imaging parameters, we assume a positional uncertainty of 5\,\ac{muas} for the robust features of the image model, which is in agreement with the width of the jet ridgelines.
Taking all constraints on the apex location into account, we estimate an uncertainty of 10\,\ac{muas} on the position.

For the determination of $z_\mathrm{core}$, the pixel and jet apex position uncertainties are added in quadrature.
Based on possible jet apex positions within the estimated uncertainty, we fit the $W \propto z^k$ jet profile multiple times and derive a systematic error of $\pm 0.06$ on $k$.
When we used image model convolved with the nominal resolving beam, we obtain $k = 0.35$ with a statistical error of $\pm 0.2$.

\subsubsection{Brightness asymmetries}
\label{sec:methods:asymm}

The jet-counterjet asymmetry is most likely caused by relativistic boosting.
We can calculate the $\mathcal{R}_\mathrm{j/cj}$ brightness ratio by taking the average image flux density within $50\times100$\,\ac{muas} rectangular regions on opposite sides of the apex.
This ratio has to be interpreted with care, since the two regions may be at different distances to the jet apex. Moreover, counterjet radiation may be absorbed by the accretion flow and intrinsic jet-counterjet differences may arise from asymmetries in the jet launching process and the ambient medium \citep{2019Baczko,2020Nathanail}.

If we assume the intrinsic emissivity to be the same in the jet sheath and spine, beaming effects can be invoked to explain observed differences in brightness across the jet. We note that the intrinsic emissivity of the jet sheath is likely larger than that of the spine, as mentioned in the main body of this work. The simplifying assumption of identical intrinsic emissivities can nonetheless be used to derive straightforward estimates for jet velocity components and the inclination angle $\theta$, since Doppler boosting is expected to have a considerable contribution to the observed source structure.
If the inclination angle $\theta$ is not too small, a significant portion of the spine emission may be beamed away from the line of sight.
If the sheath and spine velocities are $c \beta_\mathrm{sh}$ and $c \beta_\mathrm{sp}$, respectively, the ratio of $I_\mathrm{sh}$ sheath and $I_\mathrm{sp}$ spine intensities in a continuous jet follows as
\begin{equation}
\mathcal{R}_\mathrm{sh/sp} \equiv \frac{I_\mathrm{sh}}{I_\mathrm{sp}} = \left[ \frac{\sqrt{1 - \beta_\mathrm{sh}^2}\left( 1 - \beta_\mathrm{sp} \cos{\left( \theta \right)} \right)}{\sqrt{1 - \beta_\mathrm{sp}^2}\left( 1 - \beta_\mathrm{sh} \cos{\left( \theta \right)} \right)} \right]^{2-\alpha}\,,
\label{cena-eq:Rshsp}
\end{equation}
with $\alpha$~$(I \propto \nu^{\alpha})$ as the spectral index of the optically thin jet components.
Assuming a typical spectral index of $\alpha = -0.7$ and identical intrinsic emissivities, we can constrain the sheath and spine velocities with \autoref{cena-eq:Rshsp} and $\mathcal{R}_\mathrm{sh/sp} > 5$ to
\begin{equation}
\Big(1 - \beta_\mathrm{sp}^2\Big)^{-0.5}\Big( 1 - \beta_\mathrm{sp} \cos{\left( \theta \right)} \Big)
\gtrsim
1.8 \Big(1 - \beta_\mathrm{sh}^2\Big)^{-0.5}\Big( 1 - \beta_\mathrm{sh} \cos{\left( \theta \right)} \Big) \,.
\label{cena-eq:Rshsp2}
\end{equation}
$(1 - \beta^2)^{-0.5}( 1 - \beta \cos{\left( \theta \right)})$ has a minimum of $\sqrt{1-\mathrm{cos}^2{\left( \theta \right)}}$ at $\beta=\cos{\left( \theta \right)}$.
It follows that the sheath-spine asymmetry can only be explained via beaming for $\beta_\mathrm{sp} > \cos{\left( \theta \right)}$, independent of the assumed value for the spectral index.

For \ac{cena}, the jet spine emission may be beamed away from the line of sight, when its velocity exceeds $0.7\,c$\,--\,$0.9\,c$, while the sheath moves with a slower velocity. In fact, the emitting plasma of the large-scale jet was observed to move with 0.24\,c\,--\,0.37\,c \citep{Mueller2014}.

For a full three-dimensional picture of a jet, where we assume the sheath to be symmetric in the $\phi$ direction around the spine in a cylindrical coordinate system, different spine and sheath emissivities, due to beaming or intrinsic effects, cannot, on their own, explain edge-brightening.
The reason is that the sheath emission will contribute to any sightline towards the jet.
A more detailed description, where also the optical depth is taken into account, is given in the next paragraph.
In the remainder of this section, we go through the different scenarios that could cause the observed edge-brightening.
First, we discuss a common interpretation related to pathlength differences. As this only works in optically thin regions, we put the presence of helical magnetic fields forward as the most likely, intrinsic explanation for edge-brightening in \ac{llagn}.
We then discuss more exotic scenarios, of a rotating or asymmetric jet, which might be tested through future observations.

In the optically thin jet regions, the integrated column density along sightlines through the jet at different distances from its axis (center vs. edges) can be used to explain edge-brightening.
These are sightlines that, across the transverse extent of the jet, enter the jet at different locations. The sightlines first pass through the near side of the jet and exit again at the other side of the jet, the far side.
If we assume the absence of intrinsic spine emissivity (due to weak mass-loading or beaming of radiation into a narrow cone away from the line of sight), the observed radiation will be produced by a sheath of thickness $\Delta R$.
For a line sight that goes exactly through the center of the jet, we pass twice through the sheath, which would amount to a pathlength of $2 \Delta R / \sin{\theta}$, when the pathlength is short enough to locally approximate the jet as a cylinder.
For a local jet radius $R_j$, the column density along a sightline through the edge of the jet will be larger by a factor of $\sim\,\sqrt{R_j / \Delta R}$ \citep{Walker2018}. Here, we have neglected changes in emissivity as sightlines pass through material at different distances to the jet apex.
This simple model is capable of explaining edge-brightening in optically thin jet regions, where radiation along longer pathlengths accumulates.
For \ac{cena}, this would imply a thin radiating sheath with $\Delta R < 0.04 R_j$.

However, the edge-brightening in \ac{cena} extends to the presumably optically thick radio core, suggesting that different physics are at play in this jet.
The likely presence of a helical magnetic field \citep{2005Pushkarev,2012Hovatta,2018Gabuzda,2018Gabuzdab} combined with a rotating sheath and the inclination angle $\theta$, can lead to favorable/unfavorable pitch angles that maximize/minimize the synchrotron emissivity along the edges/center of the jet.
For a power-law distribution of electrons, where in the rest-frame of the jet, the electron density $n$ follows their energy $E$ as $\mathrm{d}n \propto E^{-p}\mathrm{d}E$, the synchrotron emission coefficient in the rest frame scales as $j_\nu \propto {\mid} B \sin{\chi}{\mid}^{(p+1)/2} \nu^{-(p-1)/2}$ \citep{1979Rybicki,ClausenBrown2011}. Here, $B$ is the magnetic field strength, $\chi$ is the angle between the magnetic field and line of sight, and $\nu$ is the radiation frequency.
The corresponding absorption coefficients scale as $\alpha_\nu \propto {\mid} B \sin{\chi}{\mid}^{(p+2)/2} \nu^{-(p+4)/2}$ \citep{1979Rybicki,ClausenBrown2011}.
It can be seen that no asymmetries in $\chi$ would arise across the transverse jet profile for a purely poloidal ($B_z$) magnetic field.
The edge-brightening is maximized for perpendicular angles $\chi$ between the line of sight and magnetic field at the jet edges, while the magnetic field is oriented parallel to the line of sight in the center of the jet. 
In future work, we will study the polarimetric properties of the jet with the \ac{eht} to test this hypothesis as explanation for the edge-brightening.
To get a handle on $\chi$, it will be necessary to narrow down the inclination angle $\theta$ and jet velocity with monitoring observations to take relativistic aberration into account.

For optically thick jet regions upstream of the radio core, the relativistic boosting is sensitive to the shape of the emitting region and less sensitive to the Doppler factor \citep{1979Blandford}.
In the presence of a fast helical jet flow and $\theta > 0$, part of the jet will rotate towards the observer and the other part will rotate in the opposite direction on the sky.
Beyond the initial jet launching region, the jet is strongly collimated and the viewing angle to the jet edges will be very close to $\theta$.
For a flow with toroidal and poloidal components, we denote the angle of the helical velocity component $\beta_\mathrm{h}$ with respect to the poloidal direction along the line of sight with $\phi_\mathrm{h}$.
For two identically shaped, optically thick radio core components of intensity $I_\mathrm{s}$ at the SE jet edge and $I_\mathrm{n}$ at the NW edge, we thus have \citep{Walker2018}
\begin{equation}
\mathcal{R}_\mathrm{s/n} \equiv \frac{I_\mathrm{s}}{I_\mathrm{n}} = \left[ \frac{1 - \beta_\mathrm{h} \cos{\left ( \theta + \phi_\mathrm{h} \right)}}{1 - \beta_\mathrm{h} \cos{\left ( \theta - \phi_\mathrm{h} \right)}} \right]^2  \,.
\label{cena-eq:Rsn}
\end{equation}
For a counter-clockwise jet rotation and $\mathcal{R}_\mathrm{s/n} \sim 1.6$, we get the weak constraint of
\begin{equation}
1.3 \cos{\left ( \theta - \phi_\mathrm{h} \right)} - \cos{\left ( \theta + \phi_\mathrm{h} \right)}
\simeq 0.3 \beta_\mathrm{h}^{-1} \,.
\end{equation}
When the bulk velocities of the NW and SE jet sheaths are known, $\phi_\mathrm{h}$ and subsequently $\beta_\mathrm{h}$ can be determined \citep{Walker2018}.
We note that the the small linear scales resolved by the \ac{eht} in Cen~A uniquely allow us to track relativistic dynamics across days in this source with future observations.

In an alternative scenario, this tentative NW-SE brightness asymmetry seen in \ac{cena} could be explained with two distinct jet components having different velocities or different inclinations angles with respect to the line of sight.

In this work, we have interpreted the edge-brightening in terms of a naturally emerging spine-sheath jet structure in \ac{llagn}, based on results from GRMHD simulations that are applicable to those type of sources. However, the same phenomenon is also observed in more powerful \ac{agn}; for example Cygnus~A \citep{2016Boccardia,2016Boccardib}, where an accretion flow operating at $\sim\,1\,\%$ of the Eddington limit is unlikely to be radiatively inefficient \citep{2003Tadhunter}.

\subsubsection{Collimation profile}
\label{sec:methods:col}

Following the NW and SE jet ridgelines, we bin distance values to the jet apex into intervals of 10\,\ac{muas} in size.
Within each bin, we select the brightest pixel to obtain the central location along the ridge.
We impose a statistical uncertainty of 5\,\ac{muas} on distances $z$ in accordance with the width of the jet ridgelines in our image model.
The width $W$ of the jet is taken as the distance between the two jet arms.
The profile of our image is shown in \autoref{CenA-fig:jet_fit} together with the corresponding average opening angle computed from the jet width as a function of distance to the apex.

Resolution limitations prevent us from tracing down the exact value of the initial jet opening angle $\psi_\mathrm{jet}$ near the apex, where the analysis of binned distance values becomes uncertain.
Nonetheless, we can derive an upper limit on $\psi_\mathrm{jet}$ with a simple geometric argument:
The jet has a clearly defined collimation region beyond some distance from the apex, at $z>z_\mathrm{col}$.
To estimate $z_\mathrm{col}$, we have used the SE jet arm, as it is brighter, straighter, and has a more clearly identifiable compact brightness core.
If we now assume that the jet stream converges monotonically towards the apex for $z<z_\mathrm{col}$ and that the apex itself does not correspond to an extended region, we have
\begin{equation}
\psi_\mathrm{jet} \geq 2 \arctan{\left(\frac{W\hspace{-0.1cm}\left(z_\mathrm{col}\right)}{2 z_\mathrm{col}}\right)} \,.
\label{cena-eq:Psijet}
\end{equation}

If the inclination angle $\theta$ is known, the intrinsic opening angle $\psi_\mathrm{int}$ can be computed as \citep{2017Pushkarev}
\begin{equation}
\psi_\mathrm{int} \geq 2 \arctan{\left(\sin{\left(\theta\right)} \frac{W\hspace{-0.1cm}\left(z_\mathrm{col}\right)}{2 z_\mathrm{col}}\right)}
= 2 \arctan{\left(\sin{\left(\theta\right)}\tan{\left(\frac{\psi_\mathrm{jet}}{2}\right)}\right)}
\,.
\label{cena-eq:Psiint}
\end{equation}

The jet remains collimated out to k\ac{pc} scales and contains multiple particle acceleration sites in a knotted structure \citep{Israel1998,2003Hardcastle,Mueller2011,Mueller2014}.
The source is a well-suited laboratory for models of \ac{agn} feedback \citep{1998Silk, 1998Magorrian} and the creation of ultra-high-energy cosmic rays \citep{2018Auger,2020Hess}.

\subsubsection{Confinement by the ambient medium}
\label{sec:methods:amb}

Analytic theory for axisymmetric, relativistic, poynting-dominated outflows can be used to derive exact asymptotic solutions for the influence an ambient medium on the collimation of a jet.
One can show that in the presence of external pressure gradient $P_\mathrm{ext}(z) = P_0 z^{-\kappa}$, the jet expansion profile $W$ as a function of distance along the jet axis $z$ follows \citep{2009Komissarov,2009Lyubarsky}
\begin{equation}
    \frac{d^2 W}{d z^2} - W^{-3} + C_1 P_0 z^{-\kappa} W = 0 \,,
\end{equation}
in a simplified form, with $C_1$ a numerical constant.
At large $z$ and for a shallow external pressure gradient with $\kappa < 2$, we obtain \citep{2009Komissarov,2009Lyubarsky}
\begin{equation}
    W(z) = C_1^{-1/4} \sqrt{\frac{2-\kappa}{\pi}} \sqrt{C_2^{-1} \cos^2{S} + C_2\left( C_3 \cos{S} + \frac{\pi}{2-\kappa} \sin{S} \right)} z^{\kappa/4} \,,
    \label{cena-eq:pressurecolim}
\end{equation}
for $S(z)=C_4 z^{1-\kappa/2} - C_5$ and $C_2, C_3, C_4,\,\mathrm{and}\;C_5$ numerical constants.
\autoref{cena-eq:pressurecolim} shows that the ambient pressure will confine the jet into a $W \propto z^{k}$ profile with $k=\kappa/4$.
Additionally, oscillations along the jet boundary can occur in a non-equilibrium state for $C_2 \neq (2-\kappa)/\pi$, $C_3 \neq 0$ \citep{2009Lyubarsky}.

\subsubsection{The location of the black hole}
\label{sec:methods:bhloc}

Given a measurement of the core shift $z_\mathrm{core}$ with respect to the black hole, we can gauge the observing frequency $\widetilde{\nu}$, which corresponds to a small self-absorbed nozzle region at the footprint of the jet \citep{1993Falcke}.
This region corresponds to a peak or break from a jet-dominated flat radio spectrum as it is the smallest region where particle acceleration can occur.
The minimum scale where a jet can be launched by a black hole is given by the ISCO. The size of the emission region of this nozzle would be given by the photon capture radius. Thus, we can estimate $\widetilde{\nu}$ as
\begin{equation}
\widetilde{\nu} = \frac{\nu_\mathrm{obs} \, z_\mathrm{core}\left(\nu_\mathrm{obs}\right)}{\sqrt{27} r_g}
\simeq 20 \Bigg(\frac{\nu_\mathrm{obs}}{\mathrm{GHz}}\Bigg) \Bigg(\frac{z_\mathrm{core}\left(\nu_\mathrm{obs}\right)}{\mu\mathrm{as}}\Bigg)
\Bigg(\frac{D}{\mathrm{Mpc}}\Bigg)\Bigg(\frac{M}{10^6\,M_\odot}\Bigg)^{-1}\,\mathrm{GHz}
\,.
\label{cena-eq:nutildeobs}
\end{equation}
In this expression, $M$ is the mass of the black hole and $D$ the distance from the black hole to the observer.

With the derived scaling relation of $\widetilde{\nu} \propto M^{-1} \dot{M}^{2/3} \propto M^{-1} F_{\rm r}^{8/17} D^{16/17} \, $, we can relate the break frequencies of two sources if their accretion rates or jet properties are known.
Here, $\dot{M}$ is the black hole accretion rate and $F_{\rm r}$ is the observed flat-spectrum radio flux density.
In particular, if we assume for two sources to share the same basic intrinsic jet properties and orientation with respect to Earth, we have
\begin{equation}
\frac{\widetilde{\nu_1}}{\widetilde{\nu_2}} = \left(\frac{F_{\rm r 1}}{F_{\rm r 2}}\right)^{8/17}\left(\frac{D_1}{D_2}\right)^{16/17}\left(\frac{M_1}{M_2}\right)^{-1} \,.
\label{cena-eq:nutildecomp}
\end{equation}

While these expressions are strictly speaking only true for a filled conical jet, they appear to describe the emission from the jet sheath and its basic scaling properties reasonably well \citep{2006McKinney,MoscibrodzkaFalckeNoble2016a,2019Davelaar} and allows one to make a first order estimate of the characteristic radio frequency of near-horizon emission.

We have used the above equations to estimate the accretion rate of \ac{cena} to the one of M\,87 based on the assumption of a similar coupling between \ac{smbh} inflows and jet power.
External Faraday Rotation effects and a generally variable rotation measure further complicates the assumed relation of accretion rates \citep{2021Goddi}, which should thus be taken as only an order-of-magnitude estimate.
It is however worth pointing out that the black hole growth rate measured over cosmic timescales from X-ray cavity fluxes from the jet radio lobes is $\sim 10^{-3}\,M_\odot\,\mathrm{yr}^{-1}$ for both \ac{cena} and M\,87 \citep{2006Rafferty}.

\subsection{Alternative interpretations for the brightest jet features}
\label{sec:methods:nocore}

In this work, we have interpreted the brightest jet features as radio cores, which mark the transition region between upstream synchrotron self-absorbed jet regions and downstream optically thin areas.
In our image, we are able to resolve the self-absorbed region between the putative radio core and jet apex, which coincides with the location of the \ac{smbh} and its accretion disk. With current telescopes, the radio core and upstream region remains unresolved for most \ac{agn} (see Table~2 in \cite{2017Boccardi} for example).

The radio core interpretation of the brightest jet features seems most plausible given our data.
Based on simple analytical jet theory, a bright radio core is expected to be present in \ac{vlbi} images. Radio cores are typically seen in sources similar to \ac{cena} and the core-shift typically follows the standard $\nu^{-1}$ relation in most sources \citep{2011Sokolovsky}. In fact, special circumstances have to be invoked to explain the absence of radio cores in \ac{vlbi} images. For example, obscuration by an optically thick 
region in the foreground. We do think that this is a likely scenario for our observation given the small scales probed in our image, the high observing frequency, and the proximity of the source.
Moreover, the core shift we have computed in \ac{cena} agrees with the core SED of the source and fundamental plane equations that relate the jet power of \ac{cena} to the M\,87 jet power.
Our image is dominated by the brightest, compact jet features, which would be weakly polarized and have flat spectrum as radio cores \citep{2017Boccardi}. It should be noted that wide-bandwidth \ac{aa} interferometer data, that were taken simultaneously with the \ac{vlbi} observations, show a flat spectrum between 212\,GHz and 230\,GHz and place a $3\,\sigma$ upper limit of $0.15\,\%$ on the linear polarization fraction \citep{2021Goddi}. The \ac{aa} measurements are however at a larger arcsecond resolution and we resolve out 64\,\% of the flux measured by \ac{aa} with the \ac{eht}. We therefore need future polarimetric and spectral \ac{vlbi} results for confirmation.

However, with the current observations, we cannot conclusively rule out the possibility that there is insufficient particle acceleration in the jet, such that no radio core is formed at $\lambda\,$1.3\,mm, while a core is present at longer wavelengths \citep{Mueller2011,Mueller2014}. In this scenario, the bright jet regions would most likely correspond to a shock within the jet flow.
The strongest counterargument here is that the radio spectrum of the core turns over at THz frequencies. This emission is most likely produced by the jet, ergo, particle acceleration should occur up to the energies that produce THz synchrotron emission.

\section{Data availability}

The ALMA raw visibility data can be retrieved from the \href{https://almascience.eso.org/alma-data}{ALMA data portal} under the project code 2016.1.01198.V.
The calibrated Stokes I VLBI visibility data of Centaurus A can be obtained from a DOI listed under \url{https://eventhorizontelescope.org/for-astronomers/data} with the code \texttt{2021-D03-01}.
Image FITS files and scripts to reproduce the plots are available from the corresponding author upon reasonable request.

\section{Code availability}

\sloppy

Antenna gains that enter the SEFDs were computed with \url{https://bitbucket.org/M_Janssen/eht-flux-calibration}.
The SEFDs were applied with the \url{https://github.com/sao-eht/eat} code, which also contains the EHT-HOPS pipeline.
rPICARD is hosted on \url{https://bitbucket.org/M_Janssen/picard}. Configuration and run files, which make use of self-contained Docker images are at \url{https://bitbucket.org/M_Janssen/casaeht}. This work is based on the `ER6' data production scripts, for which the {\small\texttt{30e6ca14fb50275013c668285a3b476f9bc85436\_91da63236db34f3a31b5309b18ac159128f28a35}} image was used. 

The eht-imaging software is hosted on \url{https://github.com/achael/eht-imaging}.
SYMBA is at \url{https://bitbucket.org/M_Janssen/symba}. The docker image used here is tagged as {\small\texttt{dec65699ccc0acdc6e6ba8f218d6724537fc613a}} and can be found on \url{https://hub.docker.com/r/mjanssen2308/symba}.

\newpage
\section{Supplement}

\captionsetup[figure]{labelsep=colon, name={Supplementary Figure}}

\newpage

\begin{table}[h!]
\vspace{0.5cm}
{\centering \normalsize \hspace{4.2cm} \textbf{Supplementary Table 1:} Final imaging parameters.}
\vspace{0.2cm}
\begin{center}
\begin{tabularx}{\linewidth}{@{\extracolsep{\fill}}lccc}
\hline
\hline 
\ac{lm} Gaussian                  & $\Theta_\mathrm{maj}$ & major axis of Gaussian self-calibration model & 60\,\ac{muas} \\
 self-calibration                 & $\Theta_\mathrm{min}$ & minor axis of Gaussian self-calibration model & 60\,\ac{muas} \\
                                  & $\Theta_\mathrm{PA}$  & position angle of Gaussian self-calibration model & \ang{0} \\
                                  & $S_G$ & flux density of Gaussian self-calibration model & 0.6\,\ac{jy} \\
\hline
Imaging                           & $\Phi_\mathrm{maj}$ & major axis of Gaussian image prior & 70\,\ac{muas} \\
priors                            & $\Phi_\mathrm{min}$ & minor axis of Gaussian image prior & 18\,\ac{muas} \\
                                  & $\Phi_\mathrm{PA}$  & position angle of Gaussian image prior & \ang{45} \\
                                  & $Z_0$               & flux density of Gaussian image prior & 2\,\ac{jy} \\
\hline
Regularizer                       & $\beta_z$            & weight for $Z_0$ & 0.1 \\
weights                           & $\beta_\mathrm{MEM}$ & weight for maximum entry minimization & 10 \\
\hline
Data                              & $\alpha_\mathrm{amp}^{(1,2,3)}$ & incremental weights for measured amplitudes & $0.2, 2, 10$ \\
weights                           & $\alpha_\mathrm{cp}^{(1,2,3)}$  & incremental weights for measured closure phases & $1, 10, 20$ \\
                                  & $\alpha_\mathrm{lca}^{(1,2,3)}$  & incremental weights for measured log closure amplitudes & $1, 10, 20$ \\
\hline
rPICARD image                     & $\chi^2_\mathrm{amp}$ & goodness of fit for amplitudes & 1.0 \\
goodness                          & $\chi^2_\mathrm{cp}$  & goodness of fit for closure phases & 1.8 \\
of fit                            & $\chi^2_\mathrm{lca}$ & goodness of fit for log closure amplitudes & 2.3 \\
                                  & $\mathcal{A}^\mathrm{(sc)}_\mathrm{intra}$ & mean self-calibration gain for co-located stations & 0.98 \\
                                  & $\mathcal{A}^\mathrm{(sc)}_\mathrm{\tiny{\ac{lm}}}$ & \ac{lm} mean self-calibration gain & 1.13 \\
                                  & $\mathcal{A}^\mathrm{(sc)}_\mathrm{\tiny{\ac{az}}}$ & \ac{az} mean self-calibration gain & 1.01 \\
                                  & $\mathcal{A}^\mathrm{(sc)}_\mathrm{\tiny{\ac{sp}}}$ & \ac{sp} mean self-calibration gain & 1.03 \\
\hline
EHT-HOPS image                    & $\chi^2_\mathrm{amp}$ & goodness of fit for amplitudes & 1.2 \\
goodness                          & $\chi^2_\mathrm{cp}$  & goodness of fit for closure phases & 2.1 \\
of fit                            & $\chi^2_\mathrm{lca}$ & goodness of fit for log closure amplitudes & 1.2 \\
                                  & $\mathcal{A}^\mathrm{(sc)}_\mathrm{intra}$ & mean self-calibration gain for co-located stations & 0.98 \\
                                  & $\mathcal{A}^\mathrm{(sc)}_\mathrm{\tiny{\ac{lm}}}$ & \ac{lm} mean self-calibration gain & 1.15 \\
                                  & $\mathcal{A}^\mathrm{(sc)}_\mathrm{\tiny{\ac{az}}}$ & \ac{az} mean self-calibration gain & 1.01 \\
                                  & $\mathcal{A}^\mathrm{(sc)}_\mathrm{\tiny{\ac{sp}}}$ & \ac{sp} mean self-calibration gain & 1.00 \\
\hline
\end{tabularx}
\\\vspace{0.3cm}Note -- $\mathcal{A}^\mathrm{(sc)}_\mathrm{intra}$ corresponds to the mean gain of \ac{aa}, \ac{ap}, \ac{jc}, and \ac{sm}.
\end{center}
\end{table}

\begin{figure}[h!]
\centering
\includegraphics[width=0.95\textwidth]{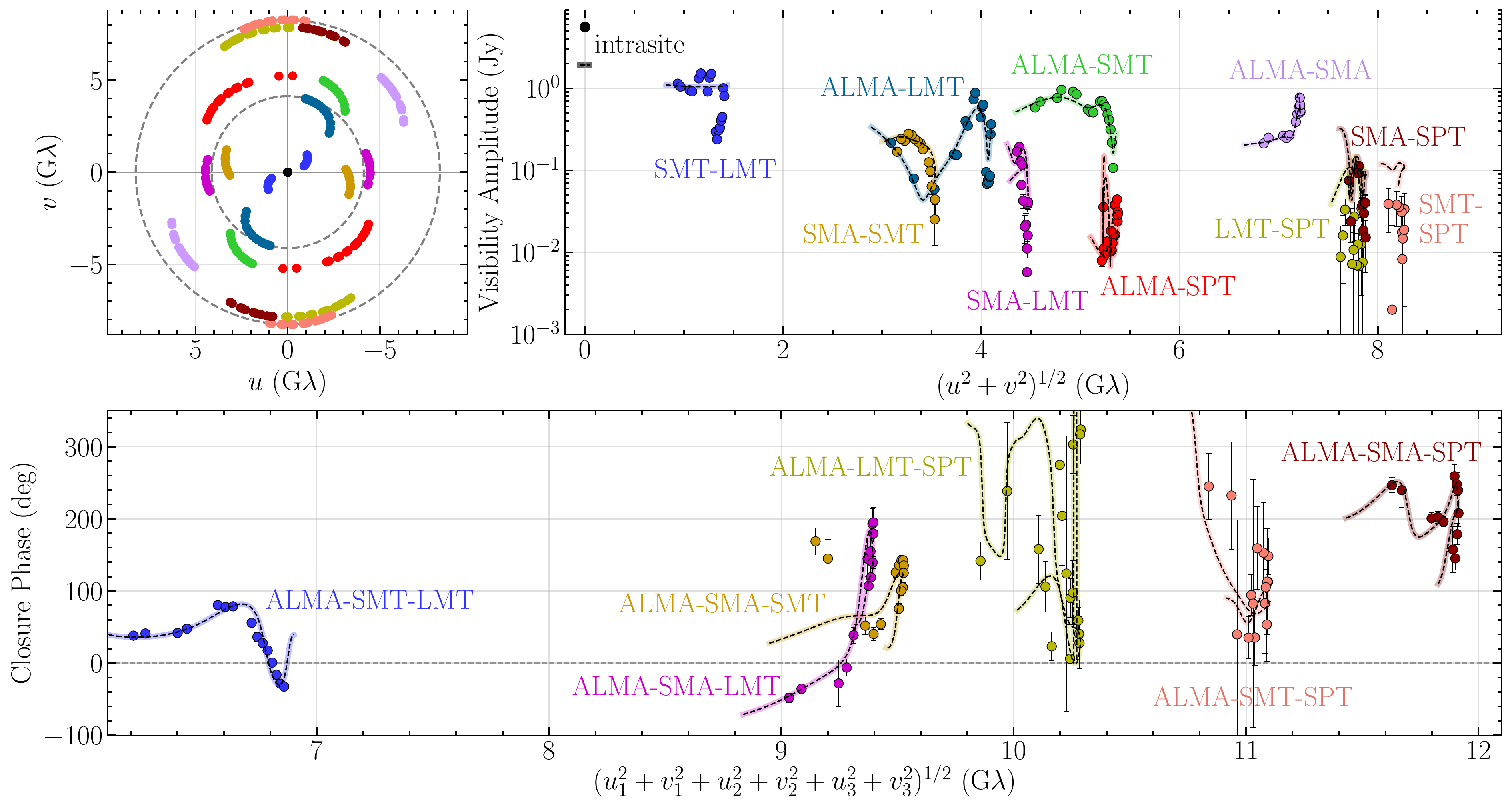}
\caption{\textbf{\ac{cena} data properties from April 2017.}
                 The top left panel shows the $(u,v)$ coverage.
                 A priori calibrated amplitude (before self-calibration) and closure phase data points are shown in the top right and bottom panel, respectively, overplotted with lines from the final image model as a function of $(u,v)$ distances. The error bars indicate thermal noise and $5\,\%$ non-closing error uncertainties added in quadrature, which are smaller than the plotted symbols in some cases. The color-coding shows different baselines. Amplitudes projected along and perpendicular to the jet position angle are given in Supplementary Fig.~\ref{CenA-SuppFIG:projplot}.
        }
\label{CenA-fig:dataplt}
\end{figure}

\begin{figure}[h!]
\centering
\includegraphics[width=0.48\textwidth]{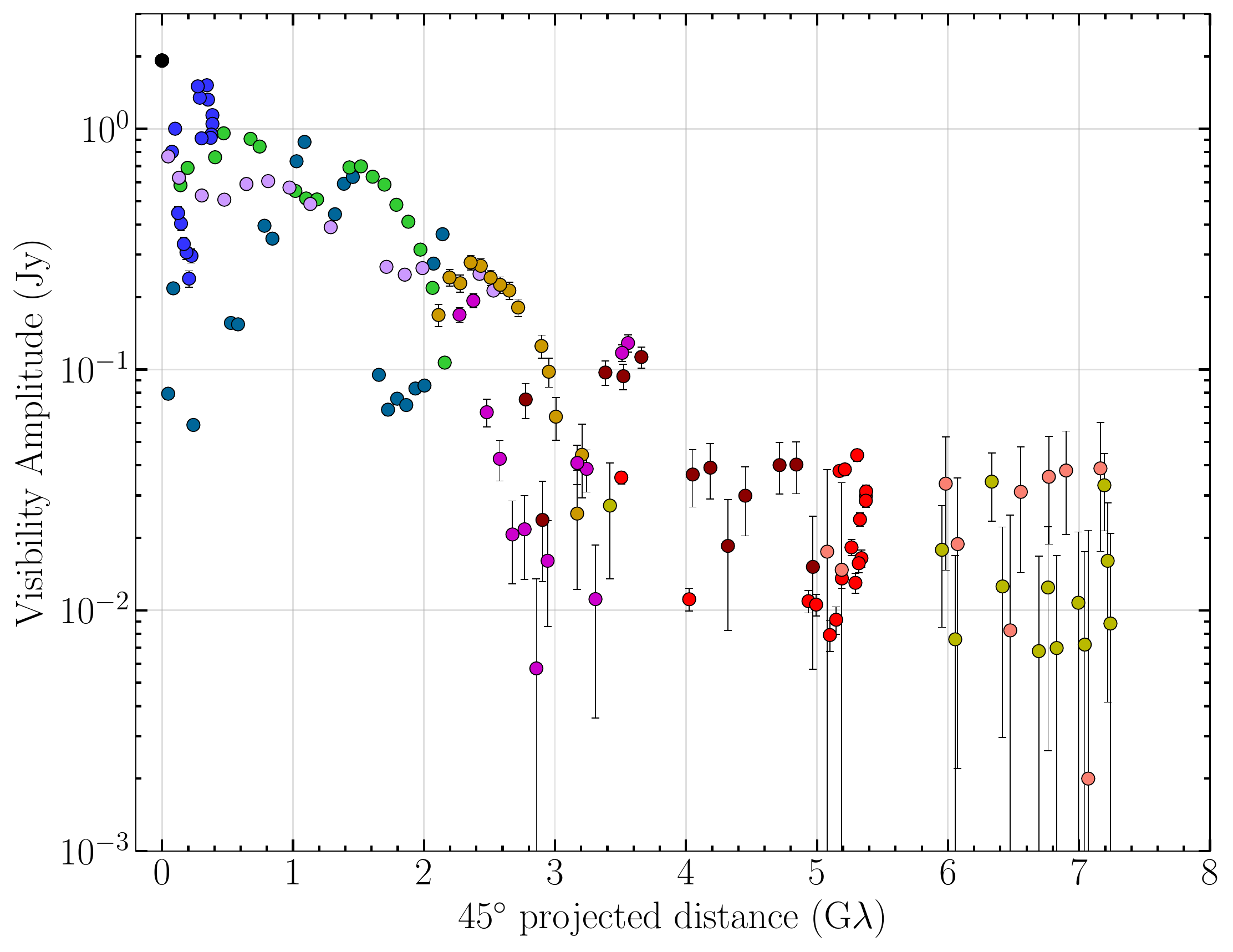}\includegraphics[width=0.48\textwidth]{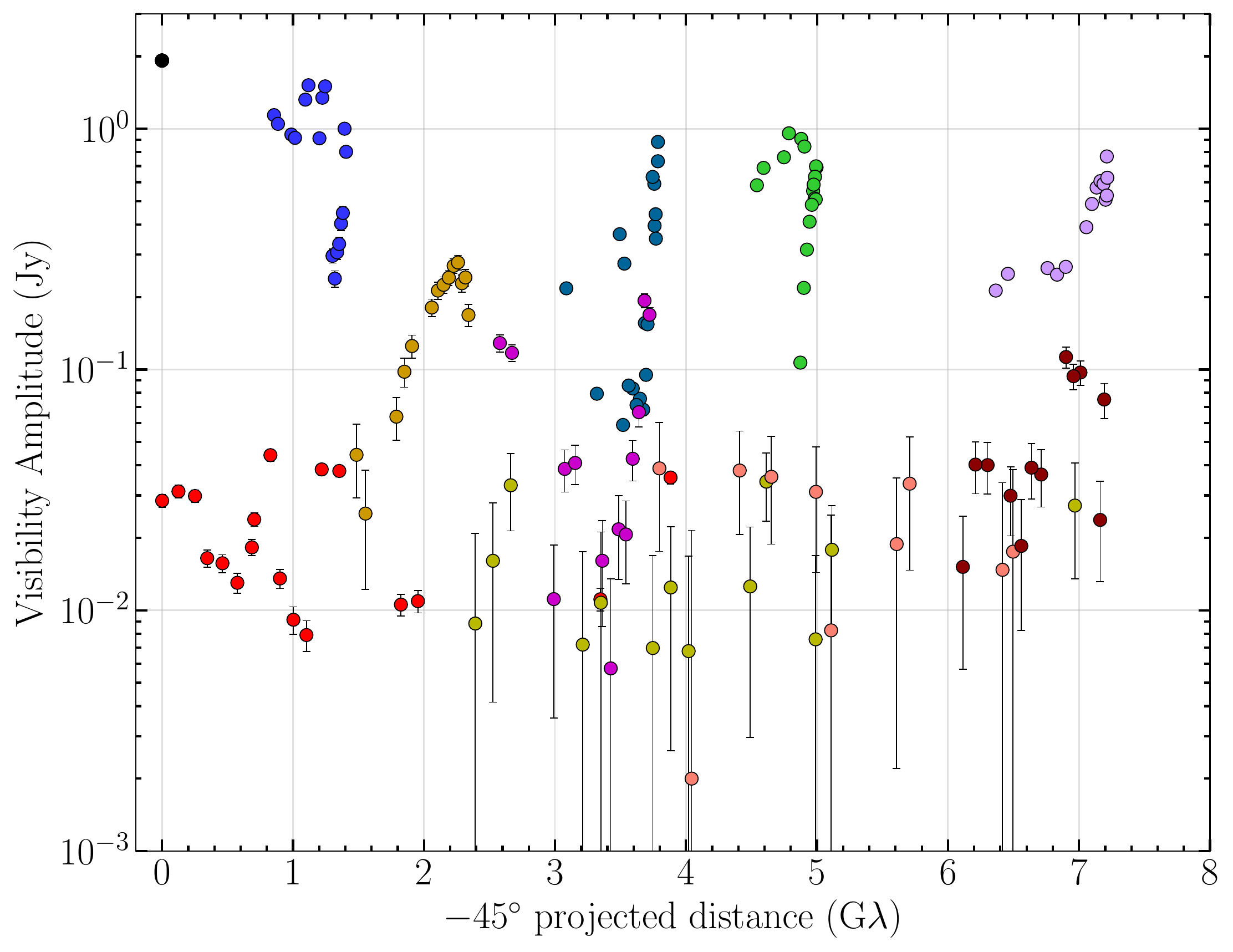}
\caption{\textbf{Source structure along specific position angles on the sky.}
                 A priori calibrated amplitudes are shown projected along the jet position angle (PA) on the sky in the left panel and perpendicular to the PA in the right panel.
                 The color coding and error bars shown are the same as in Supplementary Fig.~\ref{CenA-fig:dataplt}.}
\label{CenA-SuppFIG:projplot}
\end{figure}

\begin{figure}[h!]
\centering
\includegraphics[width=\textwidth]{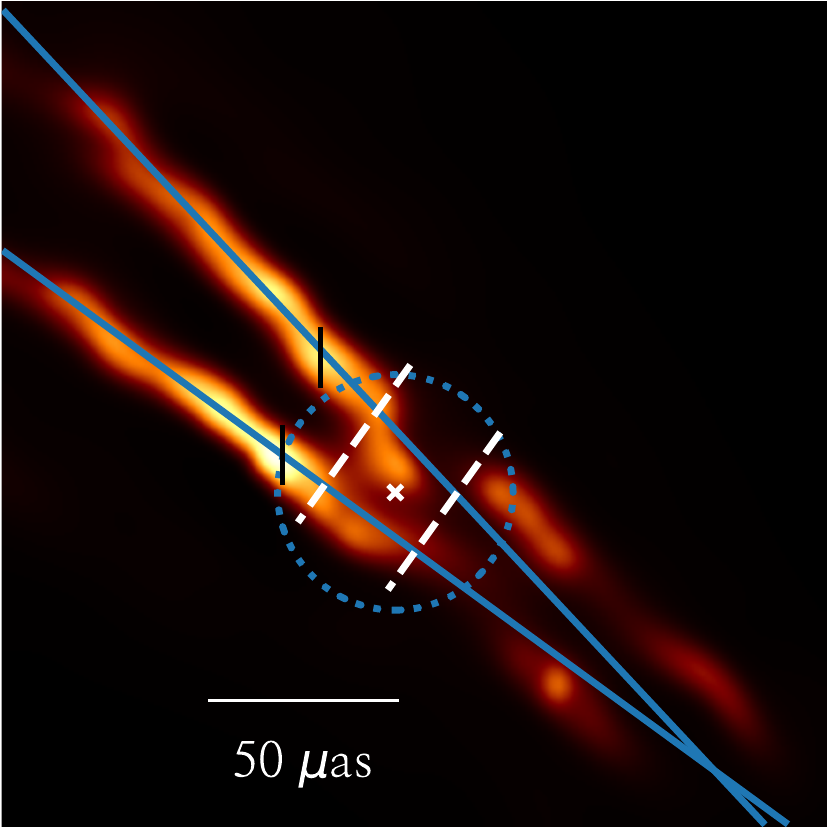}
\caption{\textbf{Determination of the jet apex location.}
                 A zoomed-in version of the final image model is shown. The solid blue lines show simple linear extrapolations of the inner NW and SE jet arms, which would place the jet apex well within the counterjet region.
                 The dashed white lines mark the certain edges of the approaching jet and the counterjet.
                 The quadrangle enclosed by the solid and dashed lines is the region where the jet apex is located.
                 Inside this quadrangle, a tentative convergence of the two streamlines can be seen.
                 The apex position assumed in this work is indicated with a white cross.
                 The surrounding blue dashed circle corresponds to the $z_\mathrm{col} = 32\,\ac{muas}$ distance.
                 Vertical black bars mark the brightest regions along each jet arm, which correspond to the assumed location of the radio core.}
\label{CenA-SuppFIG:apex-loc}
\end{figure}

\end{document}